\documentclass[12pt,preprint]{aastex}

\newcommand{\mdot}{\mbox{$\dot M$}{}}

\begin{document}
\title{Diffuse X-rays from the Arches and Quintuplet Clusters} 

\author{Gabriel Rockefeller\altaffilmark{1,2}, Christopher L.
  Fryer\altaffilmark{1,2}, Fulvio Melia\altaffilmark{1,3,4}, and Q.
  Daniel Wang\altaffilmark{5}}

\altaffiltext{1}{Department of Physics, The University of Arizona,
  1118 E 4th St, Tucson, AZ 85721}
\altaffiltext{2}{Theoretical Division, Los Alamos National Laboratory,
  P.O. Box 1663, Los Alamos, NM 87545}
\altaffiltext{3}{Steward Observatory, The University of Arizona, 933 N
  Cherry Ave, Tucson, AZ 85721}
\altaffiltext{4}{Sir Thomas Lyle Fellow and Miegunyah Fellow.}
\altaffiltext{5}{Department of Astronomy, University of Massachusetts,
  710 N Pleasant St, Amherst, MA 01003}

\begin{abstract}
  The origin and initial mass function of young stellar clusters near
  the Galactic center are still poorly understood.  Two of the more
  prominent ones, the Arches and Quintuplet clusters, may have formed
  from a shock-induced burst of star formation, given their similar
  age and proximity to each other.  Their unusual mass distribution,
  however, may be evidence of a contributing role played by other
  factors, such as stellar capture from regions outside the clusters
  themselves.  Diffuse X-ray emission from these sources provides us
  with a valuable, albeit indirect, measure of the stellar mass-loss
  rate from their constituents.  Using recent data acquired with
  \textit{Chandra}, we can study the nature and properties of the
  outflow to not only probe the pertinent physical conditions, such as
  high metallicity, the magnetic field, and so forth, but also to
  better constrain the stellar distribution within the clusters, in
  order to identify their formative history.  In this paper, we
  present a set of three-dimensional smoothed particle hydrodynamics
  simulations of the wind-wind interactions in both the Arches and
  Quintuplet clusters.  We are guided primarily by the currently known
  properties of the constituent stars, though we vary the mass-loss
  rates in order to ascertain the dependence of the measured X-ray
  flux on the assumed stellar characteristics.  Our results are
  compared with the latest observations of the Arches cluster.  Our
  analysis of the Quintuplet cluster may be used as a basis for
  comparison with future X-ray observations of this source.
\end{abstract}


\section{Introduction}

Understanding the environment's role in star formation and, in turn,
the feedback exerted by star formation on the Galactic environment, is
a problem of significance to several fields in astronomy, from the
creation of compact objects to the formation of galaxies.  The
Galactic center, with its relatively high magnetic field strength,
clouds with high particle density, and large velocity dispersions,
provides an ideal environment to study star formation under extreme
conditions \citep{M93,MF01}.  This type of investigation can benefit
from the existence of several stellar clusters in this region,
including the Arches and Quintuplet clusters.  Learning more about
their stellar constituents, and possibly their formative history, may
even give us a glimpse into the emergence of objects that will
ultimately populate the central parsec of the Galaxy
\citep[e.g.,][]{G01,M03}.

The Arches and Quintuplet clusters have been studied over a range of
wavelengths, from radio to X-ray.  X-rays provide a unique window for
investigating both the formation of binaries (point sources from
binary interactions) and the wind interactions within the entire
cluster (diffuse emission).  As \citet{Rock04} have shown, the diffuse
X-ray emission is a sensitive measure of the mass-loss rate of stars
in mutually interactive situations.  Stellar mass-loss remains one of
the largest uncertainties in stellar evolution \citep[e.g.,][]{WL99}.
X-ray observations of these clusters represent a unique probative tool
for studying the winds produced by high-metallicity systems.

In this paper, we model the propagation and interaction of winds from
stars in both the Arches and Quintuplet clusters, calculating the
X-ray fluxes arising from the consequent shocked gas.  \citet{Y02}
serendipitously discovered the Arches cluster with \textit{Chandra}
and identified five components of X-ray emission, which they labeled
A1--A5, though only A1--A3 seem to be directly associated with the
cluster \citep[see Figure 2,][]{Y02}.  A1 is apparently associated
with the core of the cluster, while A2 is located $\sim
10^{\prime\prime}$ northwest of A1.  A1 and A2 are partially resolved,
while A3 is a diffuse component that extends beyond the boundary of
the cluster; \citet{Y02} speculated that some or all of A1--A3 may be
produced by interactions of winds from stars in the system.  Analysis
of additional \textit{Chandra} observations that covered the Arches
cluster and first results from the Quintuplet cluster are presented by
\citet{W03} and \citet{L04}; these new observations resolve A1 into
two distinct components, labeled A1N and A1S, and indicate that A1N,
A1S, and A2 are all point-like X-ray sources.

Several efforts have already been made to study the X-ray
emission from clusters.  \citet{OGU97} and \citet{C00} performed
analytic calculations to estimate the diffuse emission from these
systems.  The interaction of winds in the Arches cluster has been
simulated by \citet{R01} using the \mbox{``yguaz\'{u}-a''} adaptive
grid code described in \citet{R00}.  However, all previous work
focused exclusively on the Arches cluster, and even the detailed
simulations assumed identical large values for the mass-loss rates 
($\dot M = 10^{-4} M_\odot$~yr$^{-1}$) and wind velocities ($v_{wind} 
= 1,500$~km~s$^{-1}$) of the constituent stars.  In this paper, we
present the results of simulations of both the Arches and Quintuplet
clusters, using detailed radio flux measurements (where available) and
spectral classifications to pin down the expected mass-loss rates of
stars in each system.  We then compare our results to the most recent
X-ray observations of these two clusters, including new data presented
in this paper.

A summary of relevant properties of each cluster is presented below.
We describe our numerical technique, including the characteristics of
the wind sources and the gravitational potential of the clusters, in
\S\ \ref{sect.physicalsetup} of the paper.  The new observations are
described in \S\ \ref{sect.obs}.  A comparison of the theoretical
results with the data is made in \S\ \ref{sect.results}, and the
relevance to the Galactic center conditions is discussed in \S\ 
\ref{sect.discussion}.

\subsection{The Arches Cluster}\label{sect.Arches}

The Arches stellar cluster is an exceptionally dense aggregate of
stars located at $l=0.12^\circ$, $b=0.02^\circ$, about $11^\prime$ in
projection from the Galactic center \citep[see e.g.][]{N95,C96,F02}.
The cluster is apparently a site of recent massive star formation; it
contains numerous young emission-line stars which show evidence of
strong stellar winds.  Using near-IR color-magnitudes and K band
counts, \citet{S98} estimated that at least 100 cluster members are O
stars with masses greater than $20\;M_\odot$ and calculated a total
cluster mass of $\sim(1.5$--$6)\times10^4\;M_\odot$.  \citet{F99b}
used \textit{HST} NICMOS observations to determine the slope of the
initial mass function (IMF) of the Arches cluster and calculated a
cluster mass of $\sim 10^4\;M_\odot$, with possibly 160 O stars and an
average mass density of $\sim 3\times 10^5\;M_\odot$~pc$^{-3}$.

The 14 brightest stars of this cluster have been identified with JHK
photometry and Br$\alpha$ and Br$\gamma$ hydrogen recombination lines,
showing that they have the characteristic colors and emission lines of
Of-type or Wolf-Rayet (WR) and He~\textsc{i} emission-line stars.
\citet{N95} inferred from the strength of the Br$\alpha$ and
Br$\gamma$ line fluxes that these stars are losing mass at a
prodigious rate, $\dot M\sim 2\times 10^{-5}\;M_\odot$~yr$^{-1}$, in
winds moving at $\sim 10^3$~km~s$^{-1}$.  \citet{C96} confirmed the
presence of young, massive stars using $K$-band spectroscopy; they
identified 12 stars in the cluster with spectra consistent with
late-type WN/Of objects, with mass-loss rates $\dot M\sim
(1$--$20)\times 10^{-5}\;M_\odot$~yr$^{-1}$ and wind velocities
$v_\infty\sim 800$--$1,200$~km~s$^{-1}$.

Follow-up Very Large Array (VLA) observations at centimeter
wavelengths of the brightest stars in the cluster have solidified the
detection of powerful ionized stellar winds.  Using the observed
8.5~GHz flux densities from 8 sources in the Arches cluster and the
relationship between flux density and mass-loss rate derived by
\citet{P75},
\begin{equation}\label{math-mdot}
\dot M=(5.9\times 10^{-5}\;M_\odot\,{\hbox{yr}}^{-1})\left({S_{8.5}\over
1\,\hbox{mJy}}\right)^{3/4}\left({v_\infty\over 500\;\hbox{km}\;{\hbox{s}}^{-1}}
\right)\left({d\over 8\;\hbox{kpc}}\right)^{3/2}\;,
\end{equation}
where $S_{8.5}$ is the 8.5~GHz flux density, $v_\infty$ is the wind
terminal velocity, and $d$ is the distance to the source ($\sim
8$~kpc, for stars in the Arches cluster), \citet{L01} calculated mass
loss rates $\dot M=(3$--$17)\times 10^{-5}\;M_\odot$~yr$^{-1}$,
assuming a wind electron temperature $T\sim 10^4$~K, $Z = 1$, and a
mean molecular weight $\mu = 2$.  The Wolf-Rayet phase is short-lived,
but while in this mode, stars dominate the mass ejection within the
cluster.

\begin{figure}
\epsscale{1.00}
\plotone{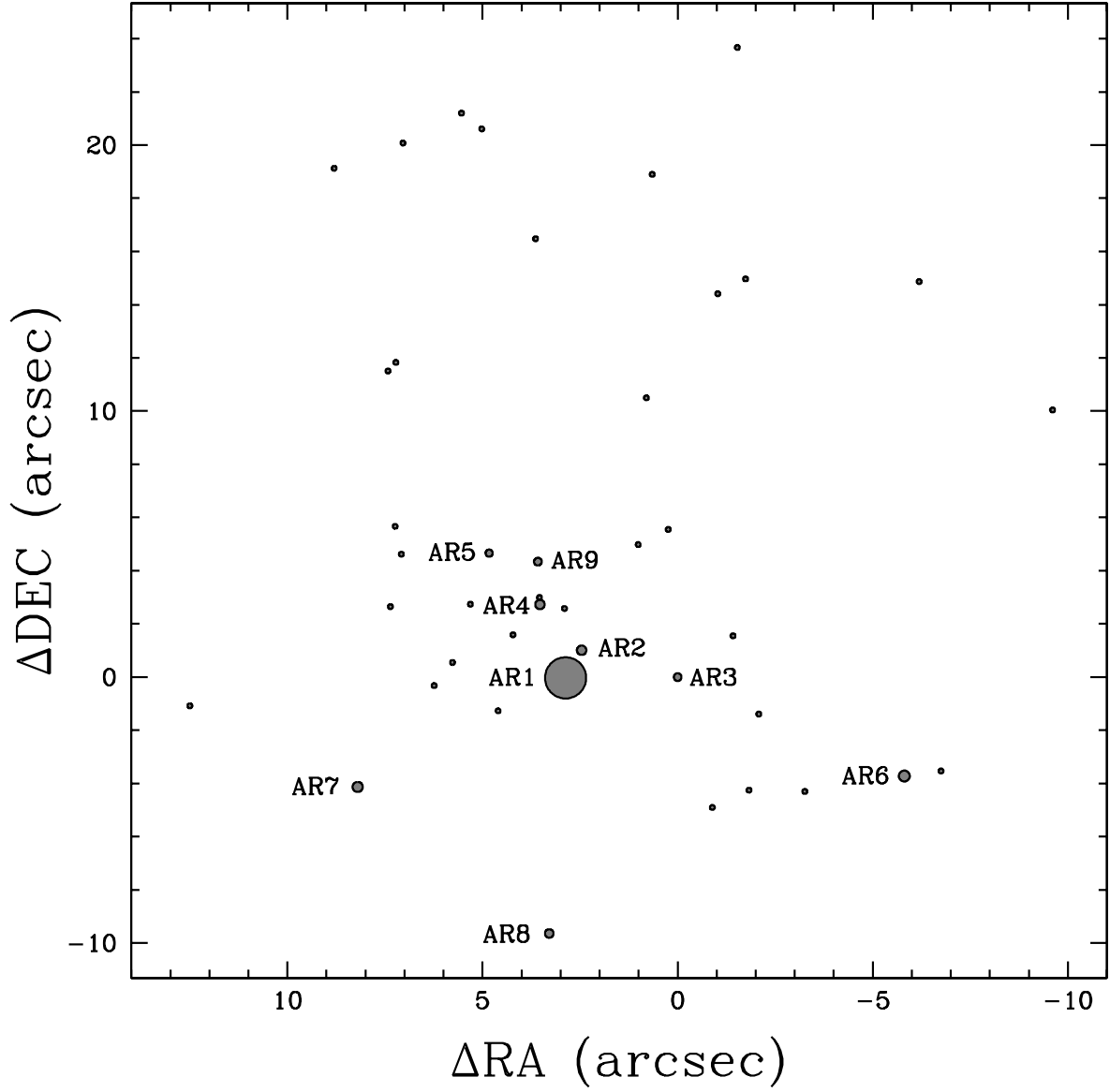}
\caption{Location of the 42 wind-producing stars used in the
  simulations of the Arches cluster, relative to $\alpha(2000)$:
  $17^h\,45^m\,50.26^s$, $\delta(2000)$:
  $-28^\circ\,49^{\prime}\,22^{\prime\prime}.76$ \citep{F02}. The radius
  of each circle corresponds (on a linear scale) to that star's mass
  loss rate.  Setting the scale is AR1, with $\dot M=1.7\times
  10^{-4}\;M_\odot$ yr$^{-1}$.}
\label{fig-arches}
\end{figure}

In an attempt to represent both the identified stellar wind sources
and the population of stars likely to be producing significant but
currently undetected winds, we include 42 wind sources (listed in
Table~\ref{Arches.srcs} and shown in Figure~\ref{fig-arches}) in our
simulations of the Arches cluster.  The stars labeled AR1--AR9
correspond to the 9 radio sources identified by \citet{L02}.  The
first 29 stars (labeled 1--29) in Table~\ref{Arches.srcs} have mass
estimates greater than $60\;M_\odot$ \citep{F02}, and are likely the
most powerful sources of wind in the cluster.  The remaining 13 stars
used in the simulations have masses less than $60\;M_\odot$ but
greater than $25\;M_\odot$ and are located on the north side of the
cluster; they are included to better represent the spatial
pattern of X-ray emission around the core of the cluster.  Based on
the broadening of the Br$\gamma$ line observed by \citet{C96}, stars
in the Arches cluster simulations are assigned wind velocities of
$1,000$~km~s$^{-1}$.  The stars labeled AR1--AR9, which have observed
8.5~GHz flux densities, are assigned mass-loss rates according to
Equation~\ref{math-mdot}.  Stars with no associated 8.5~GHz detection
but with masses larger than $60\;M_\odot$ are assigned a mass-loss
rate of $3\times 10^{-5}\;M_\odot$~yr$^{-1}$, which is equal to the
lowest mass-loss rate inferred from the weakest observed 8.5~GHz
signal from the Arches cluster \citep{L02}.  Stars with masses less
than $60\;M_\odot$ are assigned a mass-loss rate of $3\times
10^{-6}\;M_\odot$~yr$^{-1}$; their winds will have little effect on
the overall luminosity but may alter the shape of the X-ray-emitting
region.

\subsection{The Quintuplet Cluster}\label{sect.Quintuplet}

Slightly further north of Sgr A*, the Quintuplet cluster is located at
$l=0.16^\circ$, $b=0.06^\circ$.  The cluster has a total estimated
mass of $\sim 10^4\;M_\odot$ and a mass density of $\sim
10^3\;M_\odot$~pc$^{-3}$ \citep{F99a}.  Like the objects in the Arches
cluster, the known massive stars in the Quintuplet cluster have
near-IR emission-line spectra indicating that they too have evolved
away from the zero-age main sequence and now produce high-velocity
stellar winds with terminal speeds of $500$--$1,000$~km~s$^{-1}$.

\citet{F99a} obtained K-band spectra of 37 massive stars in the
Quintuplet cluster and found that 33 could be classified as WR stars,
OB supergiants, or luminous blue variables (LBVs), implying a range of
wind mass-loss rates $\dot M\sim (0.1$--$6.6)\times
10^{-5}\;M_\odot$~yr$^{-1}$.  VLA continuum images at 6~cm and 3.6~cm
of the Sickle and Pistol H~\textsc{ii} regions reveal eight point
sources located in the vicinity of the Quintuplet cluster, including
the radio source at the position of the Pistol nebula \citep{L99}.
These are labeled QR1 through 7, and the Pistol star, in
Figure~\ref{fig-quint} below.

The near-IR counterparts of QR4 and QR5 are hot, massive stars with
high mass-loss rates, one an OB~\textsc{i} supergiant and the other a
WN9/Ofpe \citep{F99a}.  The sources QR1, QR2, and QR3 also have
spectral indices consistent with stellar wind sources, but they have
no obvious NICMOS stellar counterparts.  \citet{L99} speculated that
this may be due to variable extinction associated with a dense
molecular cloud located in front of the cluster.  Given the
uncertainty, and possible variation, in stellar identification for
these 5 objects, we therefore adopt a value of $\sim 500$~km~s$^{-1}$
(typical in OB supergiants) for the speed of their wind.  The Pistol
star, on the other hand, is a prominent source in the near-IR NICMOS
image, and is evidently a luminous blue variable \citep{F98}
possessing a powerful stellar wind, though with a terminal speed of
only $v_\infty\sim 100$~km~s$^{-1}$.

In our simulations of the Quintuplet cluster, we include 31 massive
stars with spectral classifications identified by \citet{F99a}, using
estimates for the wind parameters of each star according to its
spectral type.  Wind velocities are determined according to the broad
spectral type of each star: we assume a velocity of
$1,000$~km~s$^{-1}$ for winds from WR stars, $500$~km~s$^{-1}$ winds
for OB supergiants, and $100$~km~s$^{-1}$ for the Pistol star, a LBV.

For those stars that are radio sources \citep[QR1--QR3, QR6, and QR7
from][]{L99}, mass-loss rates are determined according to
Equation~\ref{math-mdot}.  For the rest of the stars, we estimate the
mass-loss based on the spectral classifications by \citet{F99a}.  For
OB stars, the mass-loss rate was assumed to be
$10^{-5}\;M_\odot$~yr$^{-1}$ for stars with classification higher than
BO.  For smaller stars, the mass-loss rate was assumed to be
$10^{-6}\;M_\odot$~yr$^{-1}$.  For Wolf-Rayet stars, we use a
luminosity-mediated mass-loss relation \citep{WL99}:
\begin{equation}
log \left( -\dot{M}_{\rm WR}\over M_\odot\;yr^{-1} \right) = 
k+1.5 log \left(L\over L_\odot\right) - 2.85 X_s,
\end{equation}
where $L$ is the stellar luminosity from \citet{F99a}, $X_s$ is the
hydrogen mass fraction, and $k$ is a constant which we calibrated
using our radio-determined mass-loss rates.  The locations and wind
parameters of the stellar wind sources are summarized in
Table~\ref{Quint.srcs}, and their relative positions are shown in
Figure~\ref{fig-quint}.

\section{The Physical Setup}\label{sect.physicalsetup}

Our calculations use the three-dimensional smoothed particle
hydrodynamics (SPH) code discussed in \citet{FW02} and \citet{War04},
modified as described in \citet{Rock04} to include stellar wind
sources.  The gridless Lagrangean nature of SPH allows us to
concentrate spatial resolution near shocks and model gas dynamics and
wind-wind interactions on length scales that vary by several orders of
magnitude within a single calculation.

We assume that the gas behaves as an ideal gas, according to a
gamma-law ($\gamma = 5/3$) equation of state.  The effect of
self-gravity on the dynamics of the gas should be negligible compared
to the effect of the central cluster potential; we calculate
gravitational effects by approximating the potential of each cluster
with a Plummer model.

The computational domain for the simulation of each cluster is a cube
approximately $6$~pc on a side, centered on the middle of the cluster.
To simulate ``flow-out'' conditions, particles passing through the
outer boundary are removed from the simulation.  The initial
conditions assume that the space around and within each cluster is
empty; massive stars in each cluster then inject matter into the
volume of solution via winds as the calculation progresses.  The
number of particles in each simulation initially grows rapidly, but
reaches a steady number ($\sim 6.6$ million particles for simulations
of the Arches cluster and $\sim 3.3$ million particles for the
Quintuplet simulations, since there are fewer identified wind sources
in the latter) when the addition of particles from wind sources is
compensated by the loss of particles through the outer boundary of the
computational domain.

\subsection{Cluster Potential}

We model the gravitational potential $\Phi$ of a cluster using a
Plummer model \citep{P11},
\begin{equation}
\Phi(r) = \frac{-GM}{\sqrt{r^{2}+b^{2}}}\;,
\end{equation}
where $M$ is the total mass of the cluster.  The radial density
profile assumed for the cluster is therefore
\begin{equation}
\rho(r)=\rho_{0}{\left(1+\frac{r^{2}}{b^{2}}\right)}^{-5/2}\;,
\end{equation}
where the central density $\rho_{0}$ is
\begin{equation}
\rho_{0}=\frac{3M}{4\pi b^{3}}\;.
\end{equation}
We note that the spatial distributions of massive stars in the
clusters are not entirely consistent with the distribution implied by
the Plummer potential; for example, the average projected distance of
both the set of massive stars and the set of all observed stars in the
Arches cluster is roughly twice the average distance predicted by a
Plummer model based on the estimated total mass and central density of
the cluster.  We include the Plummer potential to approximate the
combined gravitational influence of the entire cluster, including the
estimated mass of stars too small or dim to be observed.

\citet{F99a} provide estimates of the total mass of each cluster by
measuring the mass of observed stars, assuming a Salpeter IMF slope,
and extrapolating down to $1\;M_\odot$---observed stars account for at
most 25\% of the mass of the Arches cluster and 16\% of the mass of
the Quintuplet cluster.  They also estimate the density of each
cluster by determining the volume of the cluster from the average
projected distance of stars from the cluster center, and dividing the
total mass by this volume; because the values are calculated using the
total cluster mass but only the average projected cluster radius, the
density estimates are probably closer to the central densities than
the average densities.  We assume that the values reported are the
central densities of the clusters.  The total cluster mass and central
density and the calculated value of $b$ for each cluster are presented
in Table~\ref{table-props}.

\subsection{Wind Sources}

We implement wind sources as literal sources of SPH particles, using
the scheme described in \citet{Rock04}.  The mass loss rates and wind
velocities inferred from observations are reported in
Tables~\ref{Arches.srcs} and \ref{Quint.srcs}.  We position each
source at its observed $x$ and $y$ location and choose the $z$
coordinate randomly, subject to the constraint that the wind sources
are uniformly distributed over a range in $z$ equal to the observed
range in $x$ and $y$.  The choice of $z$ positions has a much smaller
effect on the X-ray luminosity than the choice of mass-loss rate
(discussed below); \citet{Rock04} performed simulations of wind
sources in the central few parsecs of the Galaxy with two different
sets of $z$ positions, and the average $2-10$~keV X-ray luminosity
from the central 10$^{\prime\prime}$ of the simulations differed by
only 16\% ($7.50\times10^{31}$~erg~s$^{-1}$~arcsec$^{-2}$ from a
simulation with a dense arrangement of wind sources in the center of
the volume of solution versus
$6.45\times10^{31}$~erg~s$^{-1}$~arcsec$^{-2}$ from a simulation with
wind sources uniformly distributed in $z$).  In addition, \citet{R01}
conducted three simulations of the Arches cluster in which $z$
positions of sources were chosen by sampling from a distribution
function $f(R)\propto R^{-2}$; they found a difference in $0.5-8$~keV
X-ray luminosity of only 3\% between the most and least luminous
simulations.

Figure~\ref{fig-arches} shows the positions in the sky plane of the
wind sources in the Arches cluster, while Figure~\ref{fig-quint} shows
the wind sources in the Quintuplet cluster; the size of the circle
marking each source corresponds to the relative mass loss rate (on a
linear scale) for that star.  The initial temperature of the winds is
not well known; for simplicity, we assume that all of the winds have a
temperature of $10^{4}$~K.  Our results are fairly insensitive to this
value, however, since the temperature of the shocked gas is determined
primarily by the kinetic energy flux carried into the collision by the
winds.  The sources are assumed to be stationary over the duration of
the simulation.

We conduct two simulations of the wind-wind interactions in each
cluster; the simulations differ only in the choice of mass-loss rate
for each star, with the wind speed assumed constant.  The ``standard''
simulations use mass-loss rates inferred from observations as
described in \S\S~\ref{sect.Arches} and \ref{sect.Quintuplet}, while
the ``high-\mdot'' simulations use mass-loss rates increased from the
standard value by a factor 2.

\section{Observations}\label{sect.obs}

An on-axis \textit{Chandra} observation of the Arches cluster was
taken on 2004 June 8 for 98.6~ksec.  The ACIS-I detector was placed
at the focal plane.  The data were analyzed with the latest CIAO
(version 3.1).  While results of this observation will be presented in
an upcoming paper \citep{W04}, we concentrate here on the comparison
of the data with the simulations.  Figure~\ref{chandra-arches} shows
an overlay of an X-ray intensity contour map on a \textit{HST} NICMOS
image of the Arches cluster \citep{F99b}.  The actual spatial
resolution (with a FWHM $\lesssim 1^{\prime\prime}$) is better than
what appears on this smoothed image.  The position coincidence of the
point-like X-ray sources and bright near-IR objects is apparent.  To
compare with the simulated cluster wind properties, we remove a region
of twice the 90\% energy-encircled radius around each of the sources.

\section{Results}\label{sect.results}

All four simulations were run significantly past the point in time
when the stellar winds fill the volume of solution and gas shocked in
wind-wind collisions fills the core of each cluster.  The Arches
cluster simulations were run past $t=10,000$~yr; the Quintuplet
simulations were run for over $14,000$~yr.  The winds in each
simulation reach the edge of the volume of solution after $\sim
3,000$~yr, but the most relevant timescale for determining when the
simulations reach a steady level of X-ray luminosity is the time
required to fill the core of each cluster with shocked gas.  The
Arches cluster core is roughly five times smaller in radius than the
core of the Quintuplet cluster, so the X-ray luminosity from the
Arches simulations reaches a steady value relatively quickly compared
to the Quintuplet simulations.

\subsection{Total Flux and Time Variation}

In order to calculate the observed continuum spectrum, we assume that
the observer is positioned along the positive $z$-axis at infinity and
we sum the emission from all of the gas injected by the stellar wind
sources into the volume of each calculation.  For the conditions
encountered in the two clusters, scattering is negligible and the
optical depth is always less than unity.  For these temperatures and
densities, the dominant components of the continuum emissivity are
electron-ion ($\epsilon_{ei}$) and electron-electron ($\epsilon_{ee}$)
bremsstrahlung.  We assume that the gas is in ionization equilibrium,
although spectra obtained from the Arches cluster with
\textit{Chandra} indicate the presence of line emission at $6.4$~keV;
however, no other significant lines are observed, so our use of a
bremsstrahlung model without line emission is a reasonable
approximation.  Future, more refined versions of the calculations
reported here will need to include the effects of partial ionization
in the shocked gas, a condition suggested by the iron line emission.
We note, however, that the overall energetics and diffuse X-ray
luminosity calculated by assuming full ionization err only marginally
since the fraction of charges free to radiate via bremsstrahlung
should be very close to one.

The X-ray luminosities calculated from our simulations support the
recent result of \citet{L04} that the majority of the X-ray emission
from these clusters (e.g., $\sim 60\%$ for Arches) is probably due to
point sources.  The diffuse 0.2--10~keV X-ray luminosity calculated
from our ``standard'' Arches cluster simulation---$5.4\times
10^{34}$~erg~s$^{-1}$---falls below the 0.2--10~keV luminosity of
$4.1\times 10^{35}$~erg~s$^{-1}$ from the A1 and A2 components
reported in \citet{Y02}; the simulation using elevated estimates for
the mass-loss rates produces $2.2\times 10^{35}$~erg~s$^{-1}$ between
0.2 and 10~keV, or 53\% of the emission observed by \textit{Chandra}.
On the other hand, \citet{L04} now identify A1 and A2 as point-like
components; after subtracting the contributions of A1 and A2,
\citet{Y02} find that the A3 component has a 0.5--10~keV luminosity of
$\sim 1.6\times 10^{34}$~erg~s$^{-1}$.  Our ``standard'' simulation
produces $2.7\times10^{34}$~erg~s$^{-1}$ between 0.5 and 10~keV;
lowering the mass-loss estimates of all wind sources by $\sim 30\%$ or
slightly decreasing the assumed wind velocities would produce even
better agreement.

Our ``standard'' and ``high-\mdot'' simulations of the Quintuplet
cluster produce $1.5\times10^{33}$~erg~s$^{-1}$ and
$5.9\times10^{33}$~erg~s$^{-1}$, respectively, between 0.5 and 8~keV.
\citet{L04} estimate a 0.5--8~keV luminosity of $\sim
1\times10^{34}$~erg~s$^{-1}$ for the diffuse emission from the
Quintuplet cluster, but they point out that other regions of diffuse
emission to the north and south of the cluster introduce a complicated
pattern of background emission and limit the precision of this
estimate.

The long-term variation of the X-ray luminosity as a function of time
demonstrates some of the key differences between the two clusters.
Figures~\ref{lvst-arches} and \ref{lvst-quint} show the time variation
of the 0.5--8~keV X-ray luminosity from all four simulated clusters.
Each large graph shows the variation over the course of the entire
simulation ($> 10,000$~yr for the simulations of the Arches cluster,
and $> 14,000$~yr for the Quintuplet simulations), while each inset
graph shows variation over $1,000$ timesteps, or $\sim 150$~yr.  The
fact that the Quintuplet cluster is nearly five times larger in radius
than the Arches cluster means that much more time is required for
shocked gas to fill the central region.  The Arches cluster reaches a
steady X-ray luminosity after less than $2,000$~yr, while the
luminosity of the Quintuplet cluster does not clearly level off until
more than $10,000$~yr have passed.

The two clusters also exhibit differences in the size of short-term
fluctuations in X-ray luminosity.  Both simulations of the Arches
cluster show short-term variations in luminosity of $\sim 1\%$ over
timescales of $\sim 50$~yr, while the X-ray luminosity in the
``standard'' Quintuplet simulation fluctuates by $\sim 4\%$, and
short-term variations in the ``high-\mdot'' simulation of the
Quintuplet cluster are as large as $\sim 7\%$.  

\subsection{Spatial Variation of the X-ray Flux}

Figure~\ref{chandra-arches} shows that many of the bright X-ray peaks
in the Arches cluster correspond to actual stars, presumably binaries
whose binary wind interactions produce the strong localized X-ray
emission.  These sources must be subtracted to study the diffuse X-ray
emission.

The simulated 0.5--8~keV X-ray contours from the region near the core
of the Arches cluster, shown in Figure~\ref{contour-arches}, are
generally comparable to the contours generated from \textit{Chandra}
observations \citep[Figure~\ref{chandra-arches}; see also Figure
2b,][]{Y02}.  The strongest emission in the simulations and in
\textit{Chandra} images comes from the core of the cluster, and both
sets of contours form elliptical patterns aligned primarily
north-south.  X-ray contours from the simulation of the Quintuplet
cluster are shown in Figure~\ref{contour-quint}; the Quintuplet
cluster is significantly less dense than the Arches cluster, so the
X-ray emission is correspondingly less strongly peaked toward the
center of the cluster.

The plots in Figure~\ref{lvsr} show the total 0.5--8~keV luminosity
from concentric columns aligned along the line of sight and extending
outward in radius from the center of the Arches (left) and Quintuplet
(right) clusters.  The most luminous gas in the Arches cluster is
confined to within $20^{\prime\prime}$ of the center of the cluster;
in contrast, we include stars beyond $50^{\prime\prime}$ in the
Quintuplet cluster, and its luminosity continues to increase even
beyond a radius of $50^{\prime\prime}$.

Similarly, the plots in Figure~\ref{fvsr} show the 2--8~keV X-ray flux
per square arcmin as a function of distance from the center of each
cluster from all four simulations.  The individual crosses and error
bars in the graph from the Arches cluster represent flux measurements
from \textit{Chandra} observations of the cluster, after point sources
have been removed and an estimate of the background X-ray flux has
been subtracted.  Here we assume that all of the emission centered
around NICMOS stellar sources is due to binary wind interactions.  The
estimated background---0.064~counts~s$^{-1}$~arcmin$^{-2}$---is the
average number of counts obtained in an annulus between radii of
$0.5^{\prime}$ and $0.9^{\prime}$.  The simulations of the Arches
cluster apparently produce more X-rays near the center of the cluster
but decrease in intensity more rapidly toward larger radii.  The
relatively flat surface brightness profile evident in the
\textit{Chandra} data outside a radius of $0.3^{\prime}$ may arise in
part from the presence of additional background X-ray emission near
the Galactic center.  It may also indicate confinement of the
X-ray-emitting gas in the Arches cluster by ram pressure exerted by a
molecular cloud or other external medium surrounding the cluster; our
calculations include no such medium, so the simulated gas escapes and
cools more rapidly as it leaves the core of the cluster.

One way to constrain the amount of confining material surrounding the
cluster is to measure expansion velocities of the gas out of the
cluster.  In our simulations, which did not include confining gas, the
material accelerates until it reaches an asymptotic velocity limit
roughly equal to the mean wind velocity (Figure~\ref{vvsr}). In the
Arches cluster, we reach that limit.  In the (larger) Quintuplet
cluster, that limit apparently occurs at a distance from the cluster
center that is larger than the size of the simulation.  However, if
molecular clouds or additional stars with strong winds are producing a
confining ram pressure around this cluster, the outflowing gas will
decelerate.  Measurements of this velocity will constrain the
parameters of the surrounding material; future simulations can use
these constraints to include the effects of this material.

\section{Discussion}\label{sect.discussion}

Although the diffuse X-ray flux in clusters may be used to probe
stellar mass-loss, two notable complications in the case of the Arches
and Quintuplet clusters are the X-ray background present in the
Galactic center, and the contributions made to the overall emission by
point (i.e., binary wind) sources.  The contribution made by the X-ray
background is difficult to quantify; \citet{L04} report that
background contributions lead to significant uncertainty in the
measured X-ray flux from the Quintuplet cluster.  The contribution
from point sources is easier to handle; with its relatively high
spatial resolution, \textit{Chandra} can produce reasonable images in
which the required point-source subtraction may be made.  Point
sources in the Arches cluster all exhibit a 6.7~keV line; the absence
of such a line in the spectrum of the diffuse emission indicates that
the point-source contribution to the diffuse emission is not likely to
be significant.

If, based on the observations made by \citet{Y02}, we assume that A1
and A2 are not point-like and do not subtract the contributions of
point sources from the overall X-ray flux, it appears that the diffuse
X-ray flux is consistent with higher mass-loss rates than the
($0.3\times 10^{-5}\;M_\odot$~yr$^{-1}$) $3\times
10^{-5}\;M_\odot$~yr$^{-1}$ assumed for stars (less than) above
60\,M$_\odot$.  As we have seen, however, the observed
point-source-subtracted diffuse emission matches our calculated X-ray
flux to within a factor 2 when we adopt the currently accepted stellar
mass-loss rates.  Indeed, lowering the mass-loss rate estimates of all
wind sources by about $30\%$ would produce significant agreement
between theory and observation.  But we must make sure that we have a
complete accounting of all the wind sources and carry through with a
more careful point-source subtraction before we can completely confirm
such claims.

The fact that the simulated X-ray flux density from the Arches cluster
drops off more rapidly than the observed profile (see
Figure~\ref{fvsr}) means that (1) we may have underestimated the
contribution of the X-ray background; or, (2) we have ignored the
possibly important dynamical influence of a confining molecular gas
outside the cluster.  Interpreting data beyond these radii (roughly
$0.3^{\prime}$ for Arches and $1.0^{\prime}$ for Quintuplet) requires
more detailed information on the cluster environment there.

X-rays do prove to be an ideal probe of bulk mass-loss rates in
clusters.  The X-ray emission depends sensitively on the mass-loss
rates of the constituent stars and, based on our simulations of the
Arches cluster, we can already limit the mass-loss rates to within a
factor of 2 of the currently accepted values.  This also implies that
the assumed abundances in the Galactic center environment are
essentially correct, since the mass-loss rates from stellar models
depend on the adopted metallicity.  In addition, studying the X-ray
emission from the outer region of the cluster may eventually lead to a
better understanding of the medium within which the cluster is
embedded.  This is clearly relevant to the question of how these
unusual clusters came to be, and the relative roles played by
``standard star formation'' versus stellar capture from outside the
cluster.  The inferred stellar constituents of these clusters
\citep[e.g.,]{C96,F99a} also seem to be consistent with the required
mass loss rates, so the inferred unusual mass function of the Arches,
Quintuplet, and Central clusters continues to pose a challenge to our
understanding of how these stellar aggregates were first assembled.
 
{\bf Acknowledgments} This research was partially supported by NASA
grant NAG5-9205 and NSF grant AST-0402502 at the University of
Arizona, and has made use of NASA's Astrophysics Data System Abstract
Service.  FM is grateful to the University of Melbourne for its
support (through a Sir Thomas Lyle Fellowship and a Miegunyah
Fellowship).  This work was also funded under the auspices of the U.S.
Dept. of Energy, and supported by its contract W-7405-ENG-36 to Los
Alamos National Laboratory and by a DOE SciDAC grant number
DE-FC02-01ER41176.  The simulations were conducted primarily on the
Space Simulator at Los Alamos National Laboratory.  This research used
resources of the National Energy Research Scientific Computing Center,
which is supported by the Office of Science of the U.S. Department of
Energy under Contract No. DE-AC03-76SF00098.
{}


\clearpage
\begin{deluxetable}{lccccc}
\tablewidth{0pt}
\tablecaption{Parameters for the Arches Cluster Wind Sources\label{Arches.srcs}}
\tablehead{
  \colhead{Star\tablenotemark{a}}
& \colhead{x\tablenotemark{b}}
& \colhead{y\tablenotemark{b}}
& \colhead{z\tablenotemark{c}}
& \colhead{v}
& \colhead{\mdot} \\

& \colhead{(arcsec)}
& \colhead{(arcsec)}
& \colhead{(arcsec)}
& \colhead{(km s$^{-1}$)}
& \colhead{(${10}^{-5}\;M_\odot$ yr$^{-1}$)}
}
\startdata

1, \phn AR3 &   \phn\phs0.00 &  \phn\phs0.00 &  \phn\phs8.38 & 1,000 & \phn3.2 \\
2 &      \phn$-$6.75 &     \phn$-$3.53 & \phs10.70 & 1,000 & \phn3.0 \\
3, \phn AR7 &   \phn\phs8.20 &     \phn$-$4.13 &  \phn\phs2.66 & 1,000 & \phn4.2 \\
4, \phn AR5 &   \phn\phs4.83 &  \phn\phs4.66 &  \phn\phs2.71 & 1,000 & \phn3.0 \\
5, \phn AR8 &   \phn\phs3.29 &     \phn$-$9.64 &     \phn$-$5.31 & 1,000 & \phn3.6 \\
6, \phn AR1 &   \phn\phs2.87 &     \phn$-$0.03 &     \phn$-$1.85 &    1,000 & 17.0 \\
7, \phn AR4 &   \phn\phs3.53 &  \phn\phs2.73 &     \phn$-$5.72 & 1,000 & \phn3.9 \\
8, \phn AR2 &   \phn\phs2.46 &  \phn\phs1.01 &  \phn\phs2.78 & 1,000 & \phn3.9 \\
9 &   \phn\phs0.80 & \phs10.50 &  \phn\phs0.23 & 1,000 & \phn3.0 \\
10 &     \phn$-$1.83 &     \phn$-$4.25 &     \phn$-$0.90 & 1,000 & \phn3.0 \\
11 &     \phn$-$1.03 & \phs14.41 &  \phn\phs6.83 & 1,000 & \phn3.0 \\
12 &  \phn\phs1.01 &  \phn\phs4.98 &    $-$11.46 & 1,000 & \phn3.0 \\
13 &     \phn$-$2.08 &     \phn$-$1.39 &  \phn\phs6.12 & 1,000 & \phn3.0 \\
14 &  \phn\phs6.24 &     \phn$-$0.32 &  \phn\phs5.15 & 1,000 & \phn3.0 \\
15 &  \phn\phs7.24 &  \phn\phs5.67 &    $-$14.94 & 1,000 & \phn3.0 \\
16 &  \phn\phs4.22 &  \phn\phs1.59 &    $-$14.62 & 1,000 & \phn3.0 \\
17 &     \phn$-$0.89 &     \phn$-$4.90 & \phs14.59 & 1,000 & \phn3.0 \\
18, AR9 &  \phn\phs3.58 &  \phn\phs4.34 & \phs13.79 & 1,000 & \phn3.2 \\
19, AR6 &     \phn$-$5.81 &     \phn$-$3.72 &     \phn$-$4.97 & 1,000 & \phn4.5 \\
20 &  \phn\phs2.90 &  \phn\phs2.58 &     \phn$-$3.20 & 1,000 & \phn3.0 \\
21 &  \phn\phs7.36 &  \phn\phs2.65 & \phs10.49 & 1,000 & \phn3.0 \\
22 &  \phn\phs0.24 &  \phn\phs5.55 &     \phn$-$8.01 & 1,000 & \phn3.0 \\
23 & \phs12.50 &     \phn$-$1.08 & \phs14.43 & 1,000 & \phn3.0 \\
24 &     \phn$-$1.42 &  \phn\phs1.55 &  \phn\phs9.62 & 1,000 & \phn3.0 \\
25 &     \phn$-$3.26 &     \phn$-$4.30 &     \phn$-$8.37 & 1,000 & \phn3.0 \\
26 &  \phn\phs4.60 &     \phn$-$1.27 & \phs13.43 & 1,000 & \phn3.0 \\
27 &  \phn\phs5.31 &  \phn\phs2.74 &  \phn\phs4.80 & 1,000 & \phn3.0 \\
28 &  \phn\phs5.77 &  \phn\phs0.55 & \phs10.79 & 1,000 & \phn3.0 \\
29 &  \phn\phs7.08 &  \phn\phs4.62 & \phs11.95 & 1,000 & \phn3.0 \\
36 &     \phn$-$6.19 & \phs14.87 &     \phn$-$8.33 & 1,000 & \phn3.0 \\
37 &  \phn\phs3.54 &  \phn\phs2.99 &  \phn\phs5.89 & 1,000 & \phn3.0 \\
49 &     \phn$-$1.74 & \phs14.97 &  \phn\phs9.01 & 1,000 & \phn0.3 \\
61 &     \phn$-$1.53 & \phs23.67 &  \phn\phs2.02 & 1,000 & \phn0.3 \\
75 &  \phn\phs7.42 & \phs11.51 &  \phn\phs4.86 & 1,000 & \phn0.3 \\
108 & \phn\phs7.22 & \phs11.83 &  \phn\phs7.08 & 1,000 & \phn0.3 \\
111 & \phn\phs0.65 & \phs18.90 &    $-$12.39 & 1,000 & \phn0.3 \\
116 & \phn\phs3.64 & \phs16.48 & \phs12.31 & 1,000 & \phn0.3 \\
126 & \phn\phs8.80 & \phs19.13 &     \phn$-$0.92 & 1,000 & \phn0.3 \\
129 &    \phn$-$9.61 & \phs10.04 &  \phn\phs2.44 & 1,000 & \phn0.3 \\
132 & \phn\phs7.04 & \phs20.08 &  \phn\phs6.02 & 1,000 & \phn0.3 \\
149 & \phn\phs5.54 & \phs21.20 &     \phn$-$5.47 & 1,000 & \phn0.3 \\
156 & \phn\phs5.02 & \phs20.61 &  \phn\phs7.93 & 1,000 & \phn0.3 \\

\enddata

\tablenotetext{a}{Numerical designations taken from \citet{F02};
  ``AR'' designations taken from \citet{L02}.}

\tablenotetext{b}{Offset from $\alpha(2000)$: $17^h\,45^m\,50.26^s$,
  $\delta(2000)$: $-28^\circ\,49^{\prime}\,22^{\prime\prime}.76$
  \citep{F02}.  Here, positive $x$ is ascending R.A. (to the East) and
  positive $y$ is ascending declination (to the North).}

\tablenotetext{c}{Simulated with Monte Carlo.}

\end{deluxetable}
\clearpage


\begin{deluxetable}{lccccc}
\tablewidth{0pt}
\tablecaption{Parameters for the Quintuplet Cluster Wind Sources\label{Quint.srcs}}
\tablehead{
  \colhead{Star\tablenotemark{a}}
& \colhead{x\tablenotemark{b}}
& \colhead{y\tablenotemark{b}}
& \colhead{z\tablenotemark{c}}
& \colhead{v}
& \colhead{\mdot} \\

& \colhead{(arcsec)}
& \colhead{(arcsec)}
& \colhead{(arcsec)}
& \colhead{(km s$^{-1}$)}
& \colhead{(${10}^{-5}\;M_\odot$ yr$^{-1}$)}
}
\startdata

QR1 & \phn$-$3.30 & \phs20.10 & $-$36.41 & \phn\phm{,}500 & \phn8.0\phn \\
QR2 & \phn$-$0.30 & \phs21.80 & \phn\phs9.39 & \phn\phm{,}500 & 15.1\phn \\
QR3 & \phn\phs8.30 & \phs22.90 & $-$42.78 & \phn\phm{,}500 & \phn6.7\phn \\
QR6 & \phn$-$1.80 & \phn\phs9.00 & \phs31.80 & 1,000 & \phn6.1\phn \\
QR7 & \phn\phs3.00 & \phn\phs1.00 & $-$26.99 & 1,000 & 13.0\phn \\
76 & \phn$-$9.00 & $-$36.60 & \phn$-$9.50 & 1,000 & \phn0.28 \\
134, Pistol & \phn$-$4.50 & $-$21.90 & $-$28.26 & \phn\phm{,}100 & \phn3.8\phn \\
151 & \phn\phs1.50 & $-$18.80 & $-$20.94 & 1,000 & \phn1.1\phn \\
157 & \phs15.00 & $-$17.00 & \phn\phs8.82 & \phn\phm{,}500 & \phn0.16 \\
235 & \phn$-$4.50 & \phn\phs1.50 & \phn$-$2.91 & 1,000 & \phn3.7\phn \\
240 & $-$15.00 & \phn\phs3.80 & \phn$-$9.06 & 1,000 & \phn3.3\phn \\
241, QR5 & \phn$-$3.00 & \phn\phs4.80 & \phs39.75 & 1,000 & \phn6.6\phn \\
250 & \phn$-$7.50 & \phn\phs7.10 & \phs48.44 & \phn\phm{,}500 & \phn1.0\phn \\
256 & $-$24.00 & \phn\phs9.80 & \phn$-$9.66 & 1,000 & \phn0.70 \\
257 & \phn$-$4.50 & \phn\phs9.50 & $-$19.34 & \phn\phm{,}500 & \phn1.0\phn \\
269 & \phn$-$9.00 & \phs11.90 & \phs24.33 & \phn\phm{,}500 & \phn0.1\phn \\
270, QR4 & \phn$-$3.00 & \phs13.10 & $-$29.77 & \phn\phm{,}500 & \phn1.4\phn \\
274 & $-$39.00 & \phs12.60 & \phs42.80 & 1,000 & \phn0.67 \\
276 & \phs22.50 & \phs12.70 & \phn$-$7.87 & \phn\phm{,}500 & \phn0.1\phn \\
278 & \phn$-$3.00 & \phn\phs7.50 & $-$36.93 & \phn\phm{,}500 & \phn1.0\phn \\
301 & $-$16.50 & \phs20.00 & \phn$-$4.93 & \phn\phm{,}500 & \phn0.1\phn \\
307 & \phn$-$9.00 & \phs21.50 & $-$41.15 & \phn\phm{,}500 & \phn1.0\phn \\
309 & $-$39.00 & \phs22.90 & $-$38.32 & 1,000 & \phn0.26 \\
311 & \phs18.00 & \phs22.50 & \phs34.69 & \phn\phm{,}500 & \phn0.1\phn \\
320 & \phs12.00 & \phs25.30 & $-$18.66 & 1,000 & \phn0.61 \\
344 & $-$27.00 & \phs32.40 & \phn\phs4.04 & \phn\phm{,}500 & \phn0.1\phn \\
353 & \phs55.50 & \phs36.70 & \phn$-$6.15 & 1,000 & \phn0.22 \\
358 & $-$25.50 & \phs36.80 & $-$15.30 & \phn\phm{,}500 & \phn0.32 \\
362 & $-$46.50 & \phs38.40 & $-$13.75 & \phn\phm{,}100 & \phn3.0\phn \\
381 & \phs21.00 & \phs42.80 & $-$22.20 & \phn\phm{,}500 & \phn0.40 \\
406 & \phs15.00 & \phs51.50 & $-$26.93 & \phn\phm{,}500 & \phn0.1\phn \\

\enddata

\tablenotetext{a}{Numerical designations taken from \citet{F99a};
  ``QR'' designations taken from \citet{L02}.}

\tablenotetext{b}{Offset from $\alpha(1950)$: $17^h\,43^m\,4.5^s$,
  $\delta(1950)$: $-28^\circ\,48^{\prime}\,35^{\prime\prime}$
  \citep[based on][]{L99}. Here, positive $x$ is ascending R.A. (to
  the East) and positive $y$ is ascending declination (to the North).}

\tablenotetext{c}{Simulated with Monte Carlo.}

\end{deluxetable}
\clearpage


\begin{deluxetable}{lcccc}
\tablewidth{0pt}
\tablecaption{Properties of the Arches and Quintuplet Clusters\label{table-props}}
\tablehead{
  \colhead{Cluster}
& \colhead{M\tablenotemark{a}}
& \colhead{$\rho_{0}$\tablenotemark{a}}
& \colhead{b} \\

& \colhead{($M_{\odot}$)}
& \colhead{($M_{\odot}$ pc$^{-3}$)}
& \colhead{(pc)}
}
\startdata

Arches      & $2.0\times 10^{4}$ & $6.3\times 10^{5}$ & 0.20 \\
Quintuplet  & $6.3\times 10^{3}$ & $1.6\times 10^{3}$ & 0.98 \\

\enddata

\tablenotetext{a}{See \citet{F99a}.}
\end{deluxetable}
\clearpage
\clearpage
\begin{figure}
\epsscale{1.00}
\plotone{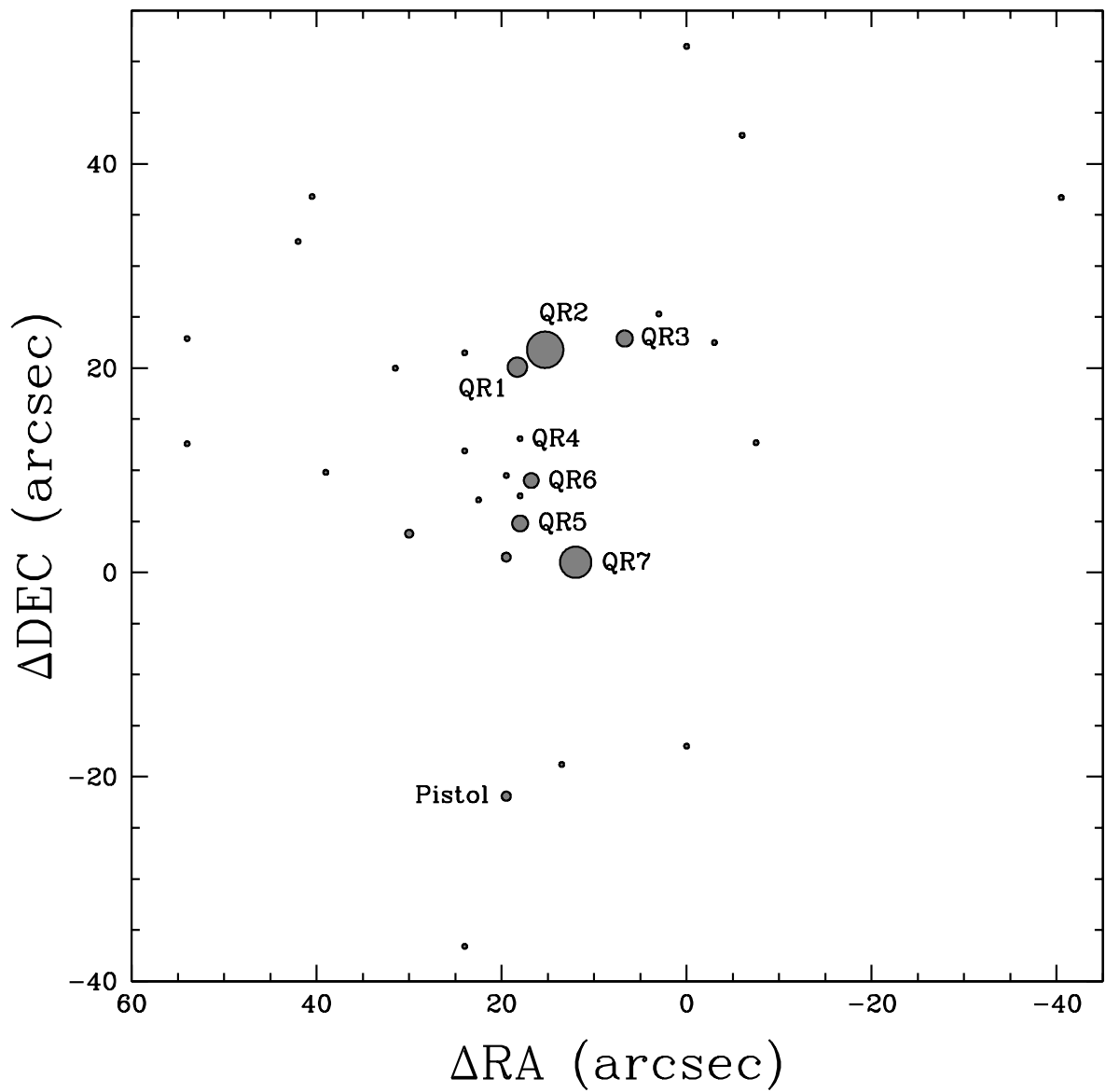}
\caption{Location of the 31 wind-producing stars used in the
  simulations of the Quintuplet cluster, relative to $\alpha(1950)$:
  $17^h\,43^m\,4.5^s$, $\delta(1950)$:
  $-28^\circ\,48^{\prime}\,35^{\prime\prime}$ \citep{L99}. The radius
  of each circle corresponds (on a linear scale) to that star's mass
  loss rate.  Setting the scale is QR2, with $\dot M=1.5\times
  10^{-4}\;M_\odot$ yr$^{-1}$.}
\label{fig-quint}
\end{figure}
\clearpage
\begin{figure}
\epsscale{1.00}
\includegraphics[angle=90]{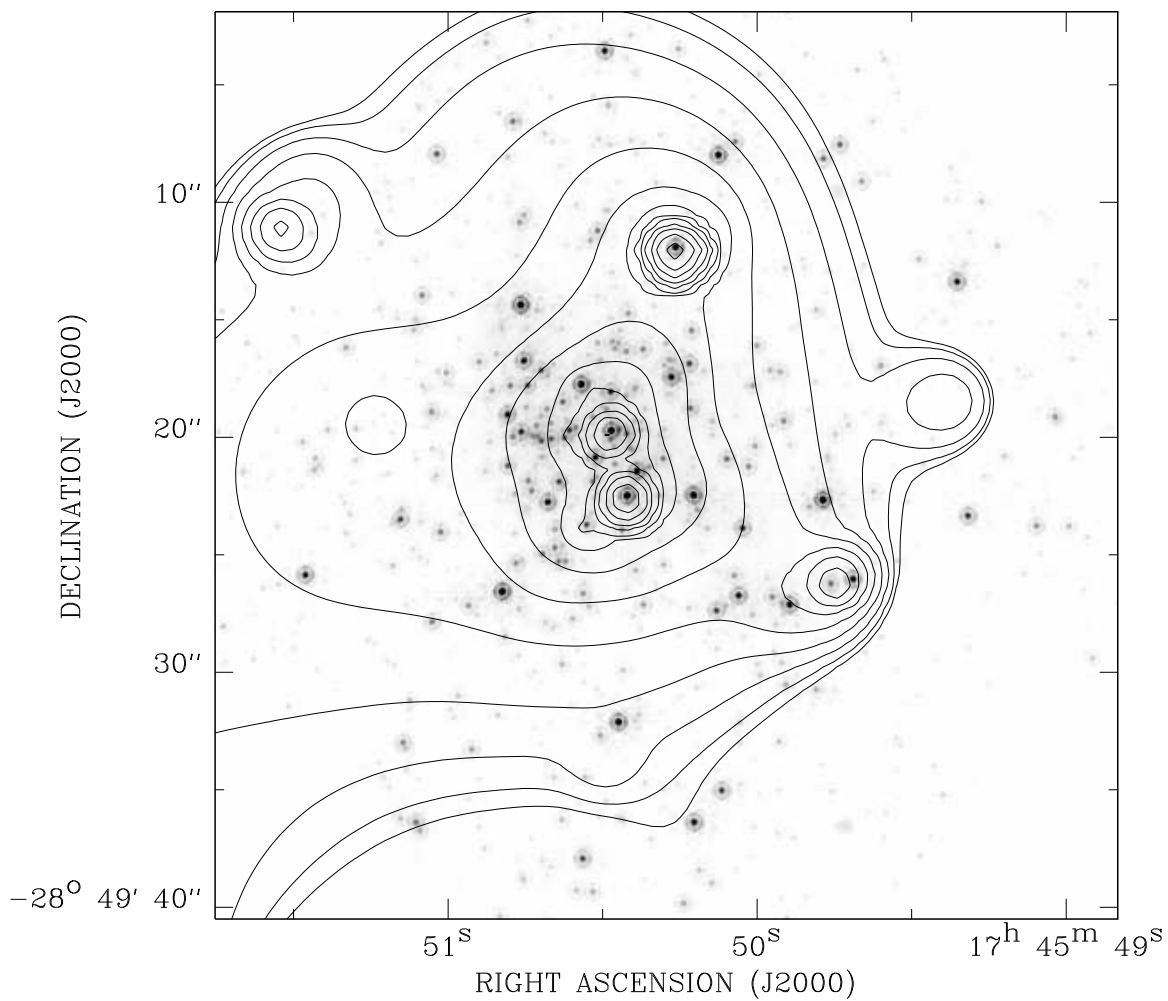}
\caption{\textit{Chandra} ACIS-I 1--9 keV intensity contours overlaid on a 
  \textit{HST} NICMOS image of the Arches cluster. The image is
  exposure-corrected and is adaptively smoothed with the CIAO csmooth
  routine ($S/N \sim 3\sigma$).  The contour levels are at $(31, 32,
  34, 38, 46, 62, 94, 158, 286, 542, 1054, 2078, 4126) \times 10^{-3}
  {\rm~counts~s^{-1}~arcmin^{-2}}$.}
\label{chandra-arches}
\end{figure}
\clearpage
\begin{figure}
\epsscale{1.00}
\plottwo{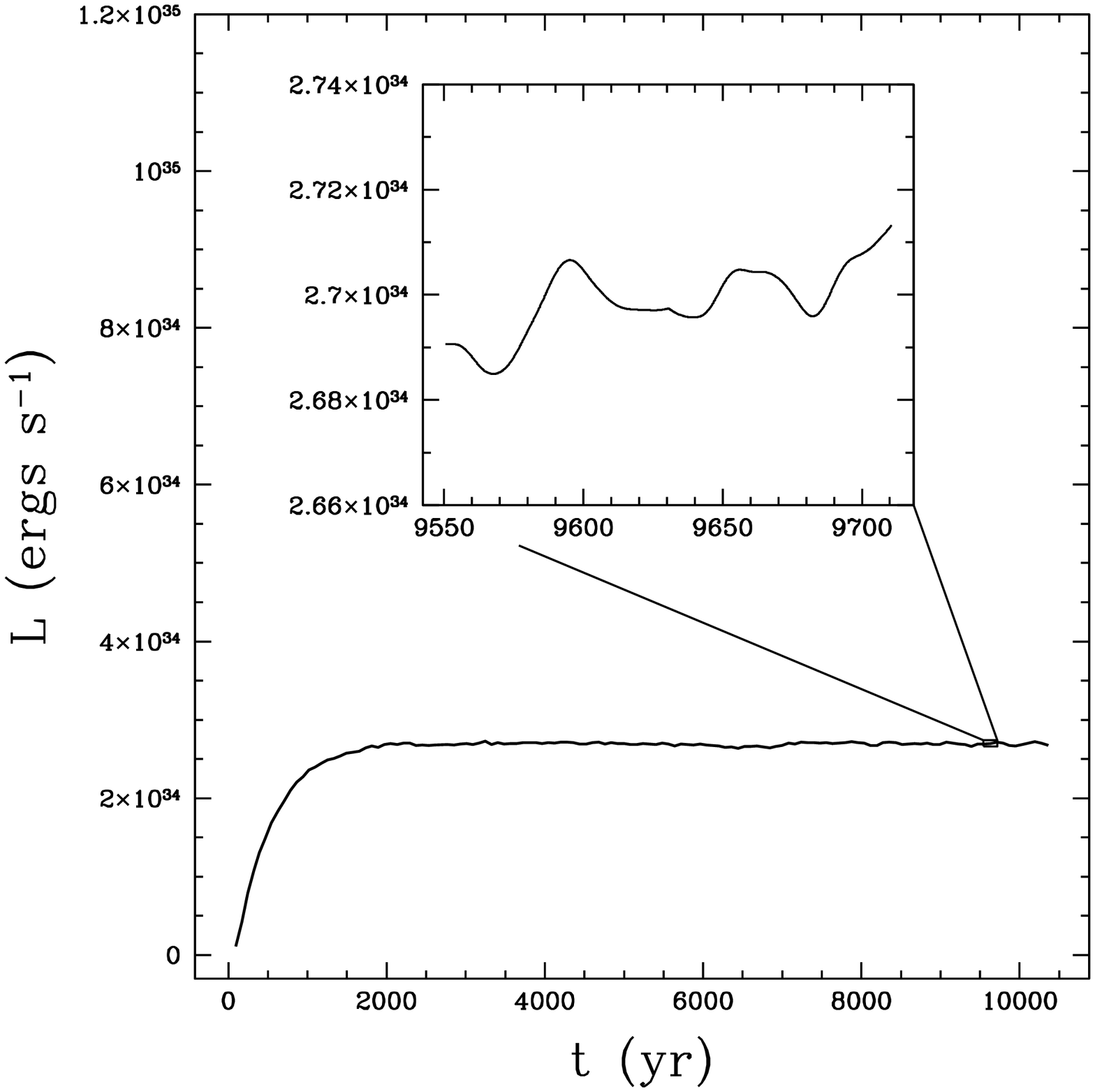}{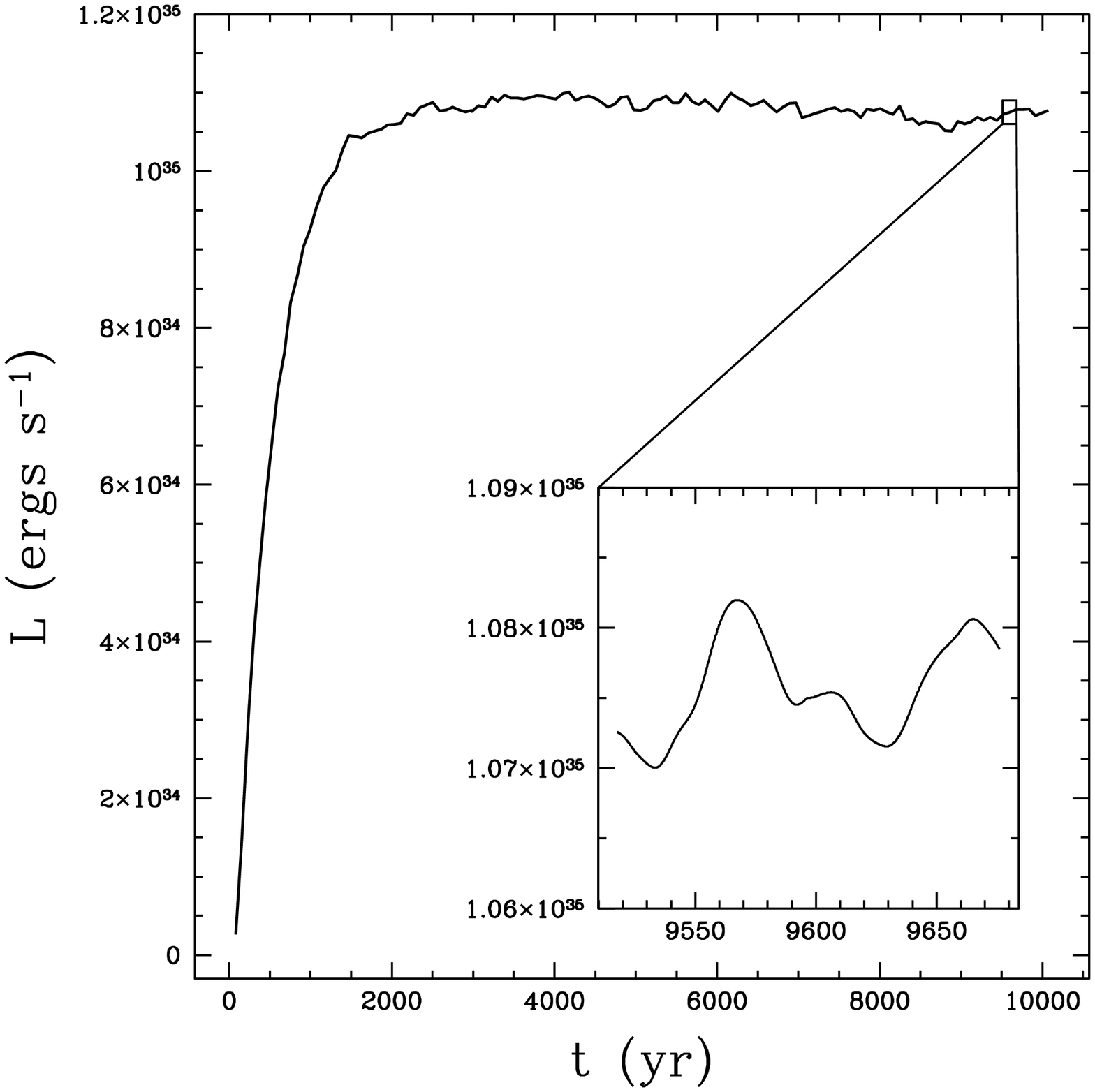}
\caption{The 0.5--8 keV X-ray luminosity versus time from the ``standard'' 
  (left) and ``high-\mdot'' simulations of the Arches cluster.  The
  large plot shows the variation in luminosity over the entire
  calculation, while the inset plot shows variation on a timescale of
  $\sim$~10~years.  The winds fill the core of the cluster after
  $\sim$~2,000 years.}
\label{lvst-arches}
\end{figure}
\clearpage
\begin{figure}
\epsscale{1.00}
\plottwo{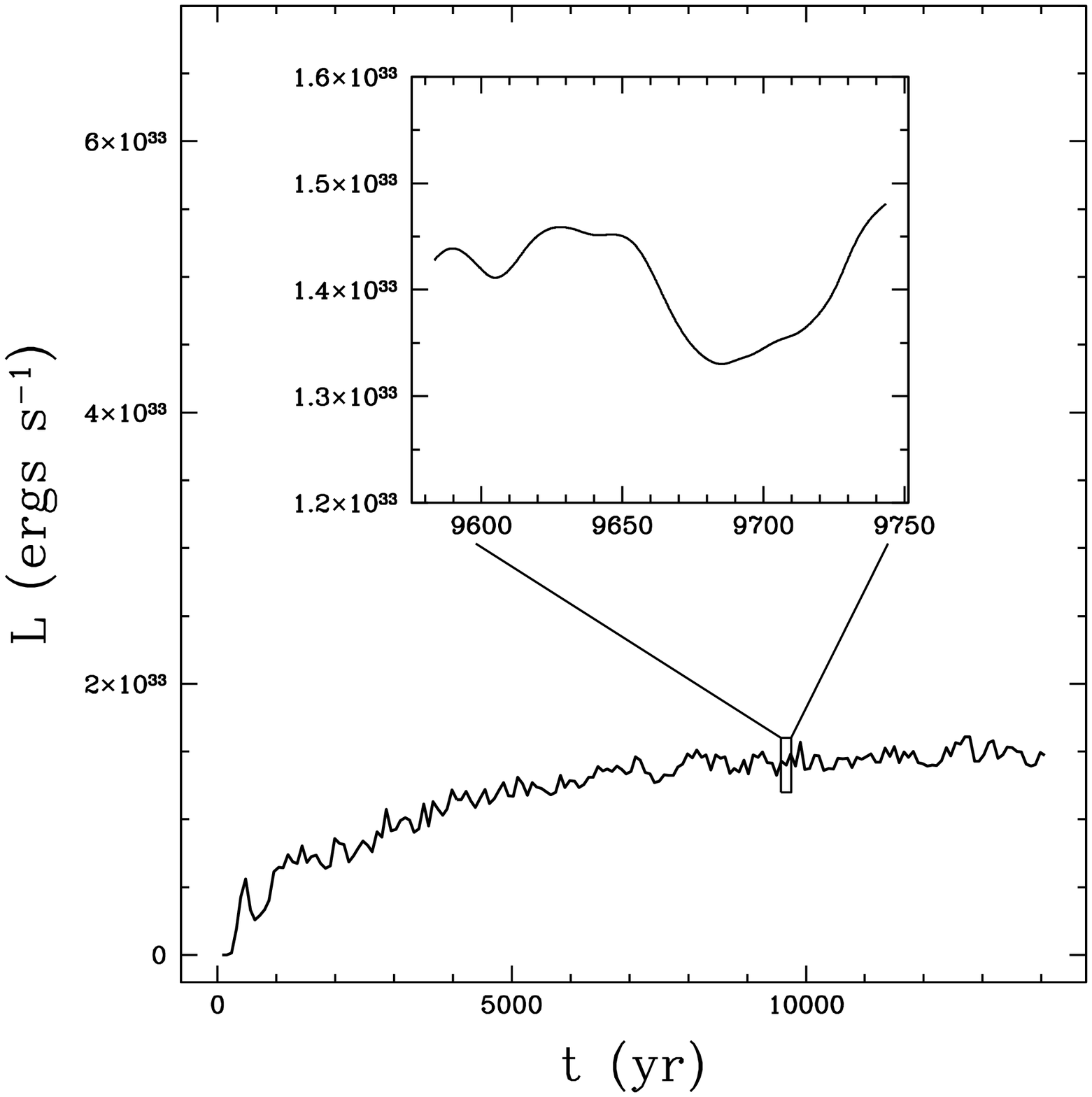}{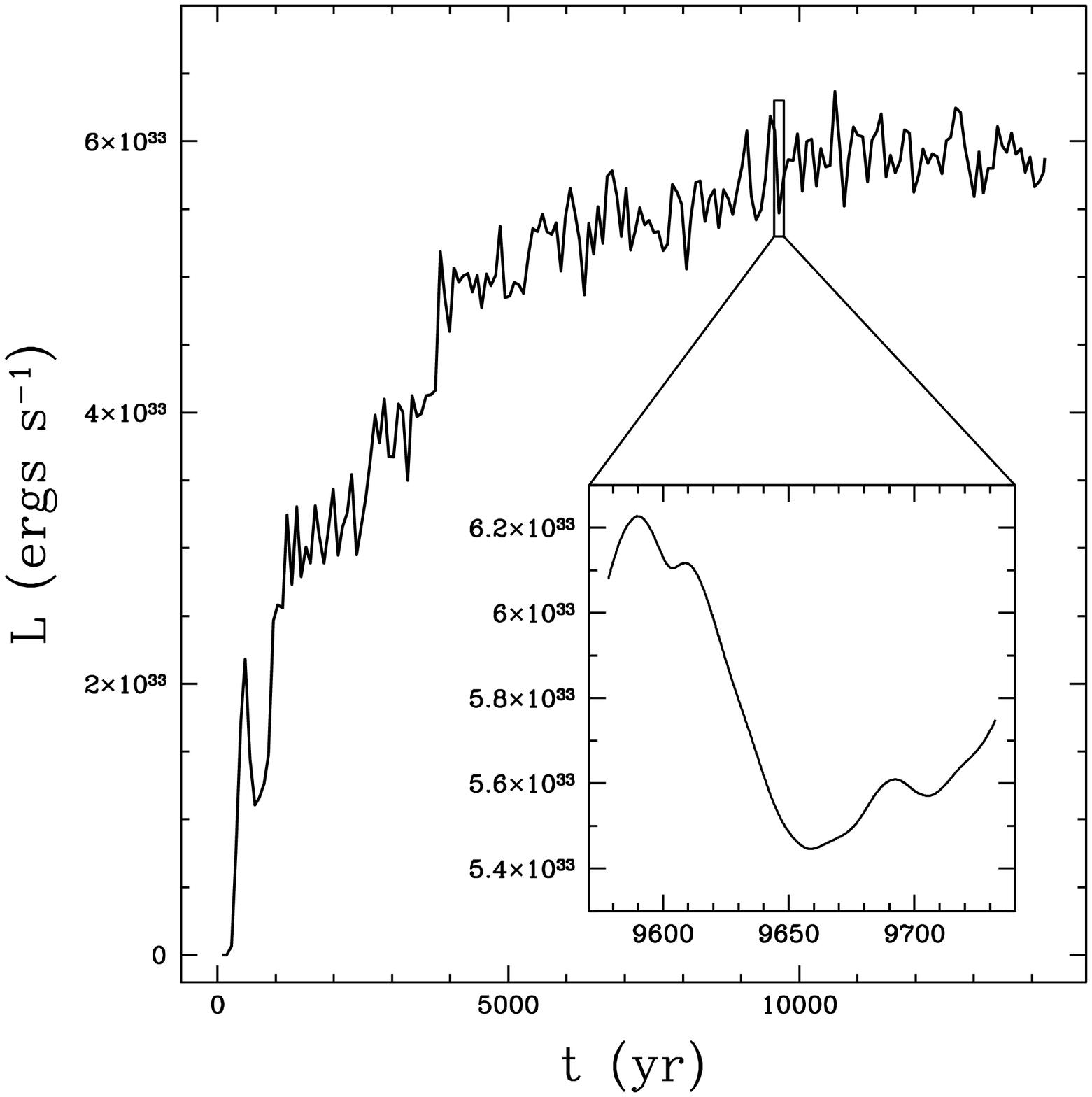}
\caption{The 0.5--8 keV X-ray luminosity versus time from the ``standard'' 
  (left) and ``high-\mdot'' simulations of the Quintuplet cluster.
  The large plot shows the variation in luminosity over the entire
  calculation, while the inset plot shows variation on a timescale of
  $\sim$~10~years.  The winds fill the core of the cluster after
  $\sim$~10,000 years.}
\label{lvst-quint}
\end{figure}
\clearpage
\begin{figure}
\epsscale{1.00}
\plottwo{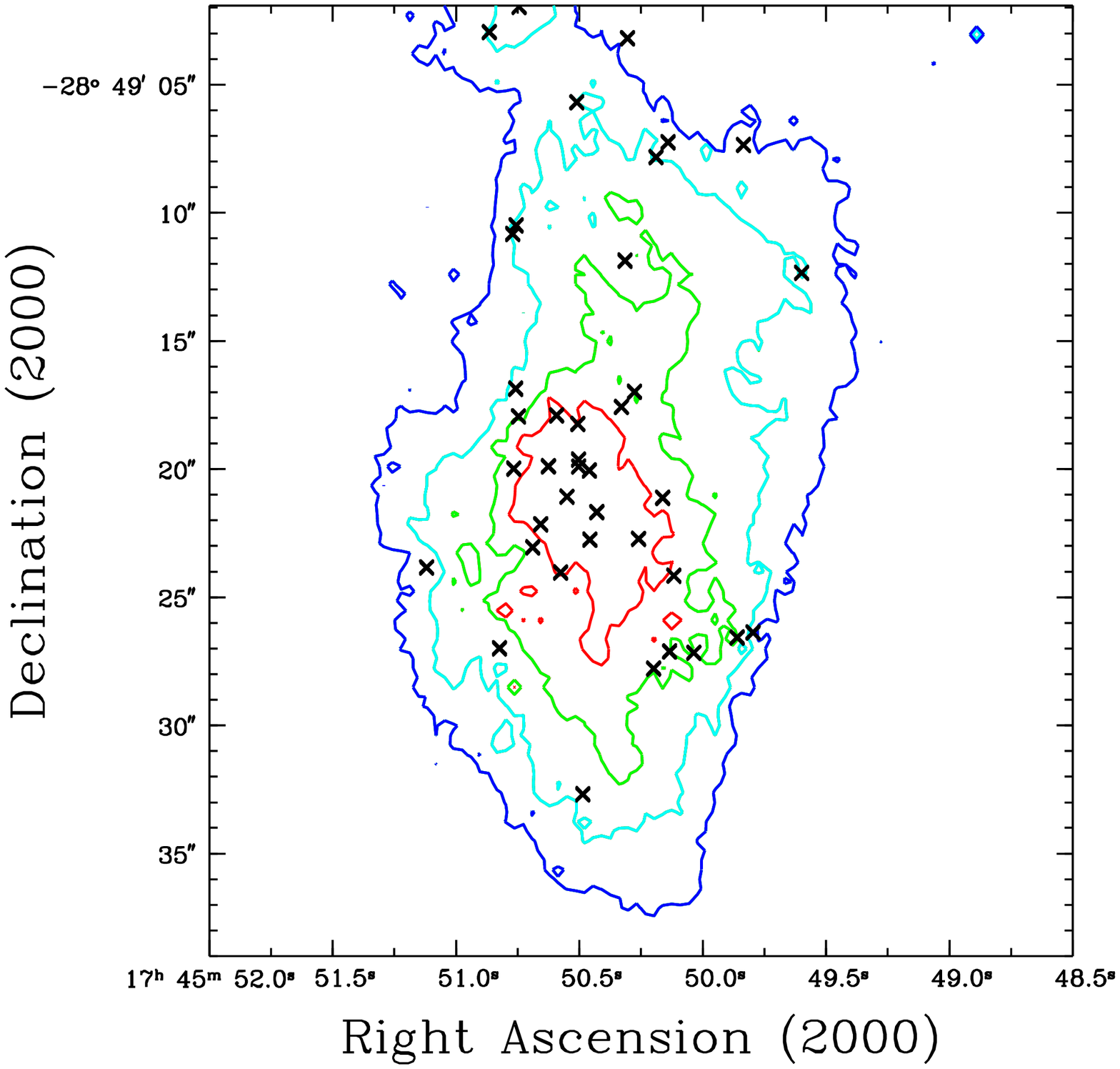}{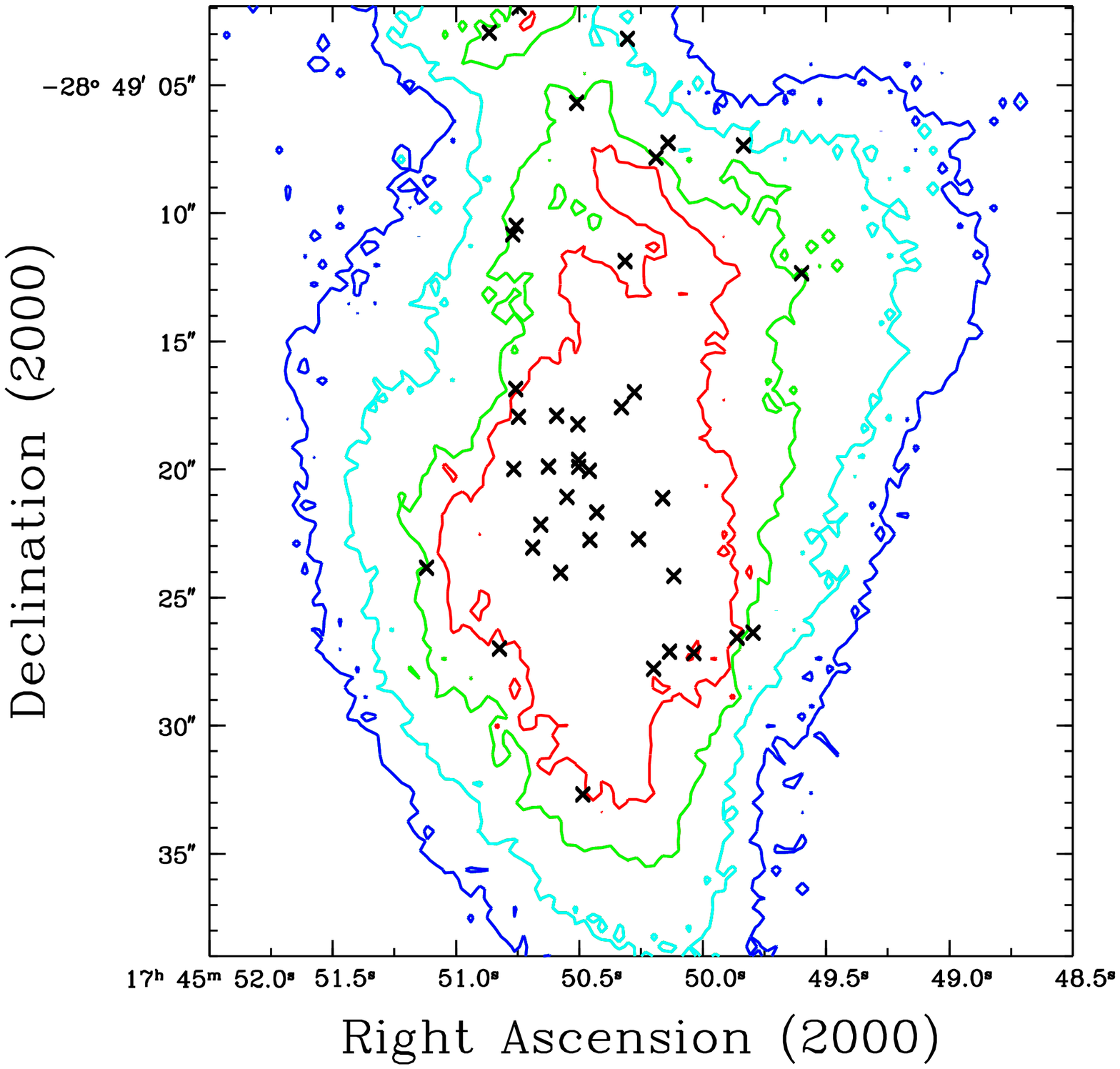}
\caption{Contours of column-integrated 0.5--8 keV X-ray luminosity per
  arcsec$^2$ from the ``standard'' (left) and ``high-\mdot''
  simulations of the Arches cluster.  In order from blue to cyan to
  green to red, the luminosities indicated by the contours are
  $10^{30}$, $2.5\times10^{30}$, $8\times10^{30}$, and
  $2\times10^{31}$~ergs~s$^{-1}$~arcsec$^{-2}$.  The crosses indicate
  the positions of wind sources included in the simulations.}
\label{contour-arches}
\end{figure}
\clearpage
\begin{figure}
\epsscale{1.00}
\plottwo{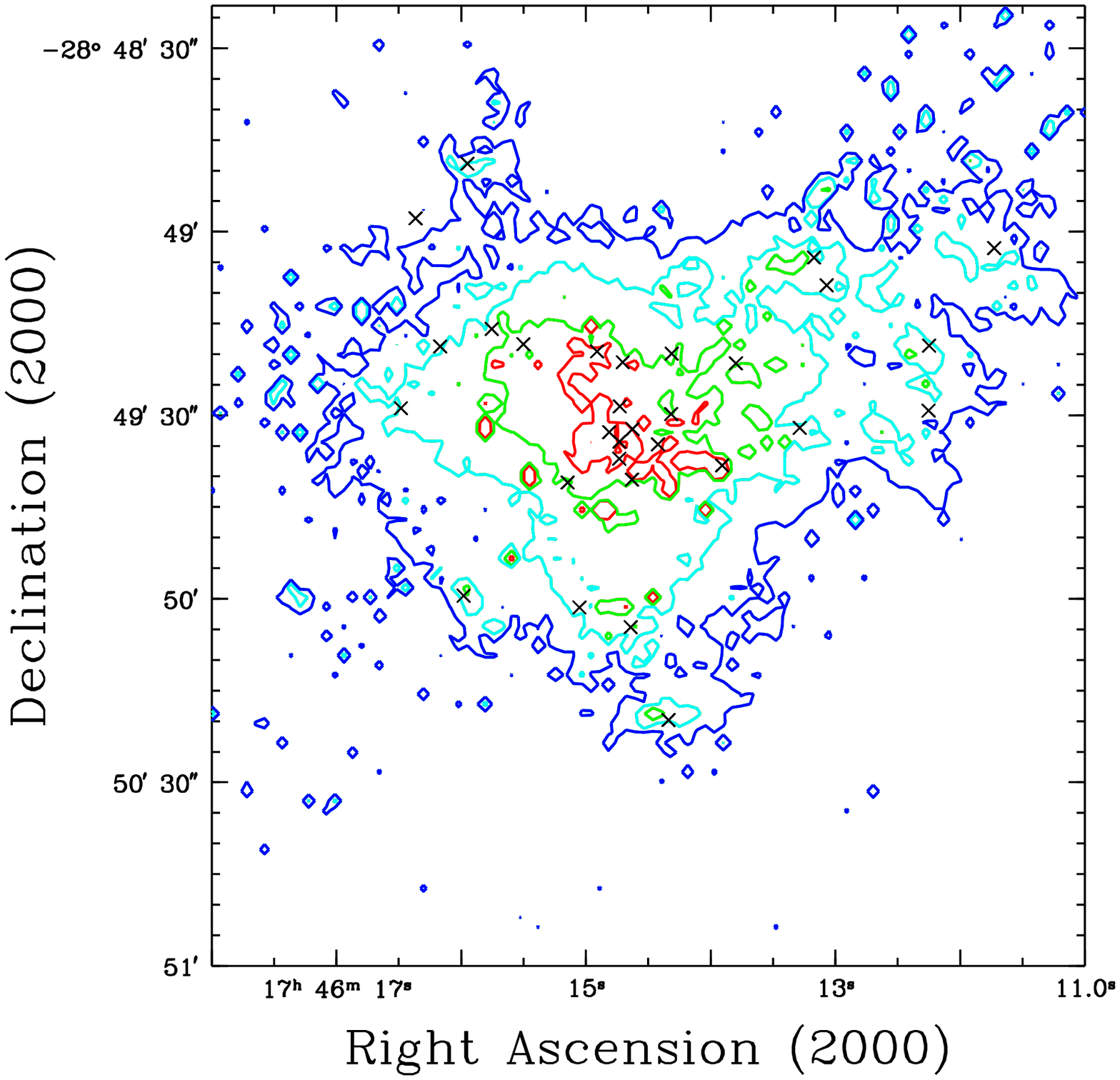}{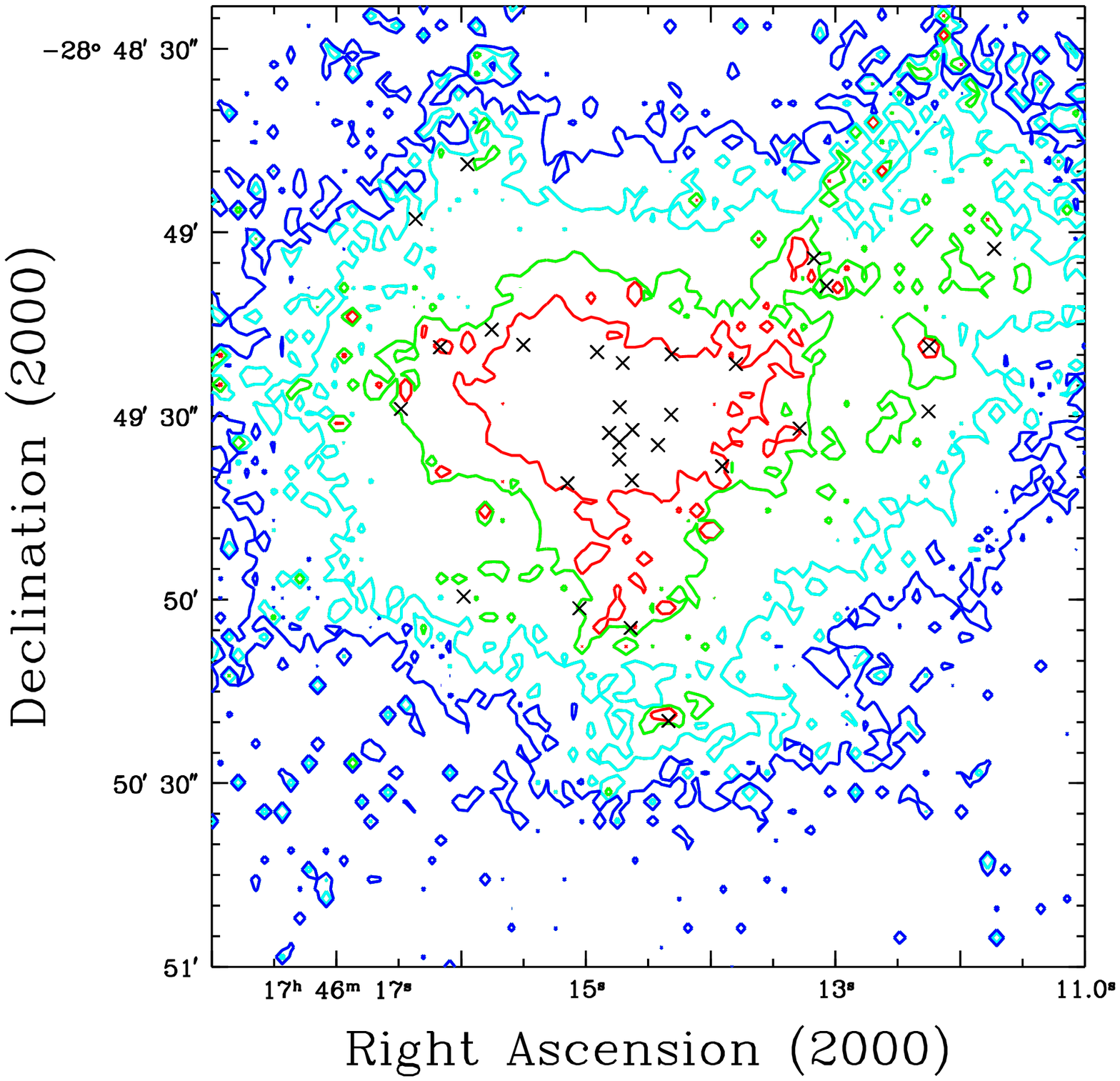}
\caption{Contours of column-integrated 0.5--8 keV X-ray luminosity per
  arcsec$^2$ from the ``standard'' (left) and ``high-\mdot''
  simulations of the Quintuplet cluster.  In order from blue to cyan
  to green to red, the luminosities indicated by the contours are
  $7\times10^{28}$, $1.8\times10^{29}$, $7\times10^{29}$, and
  $2\times10^{30}$~ergs~s$^{-1}$~arcsec$^{-2}$.  The crosses indicate
  the positions of wind sources included in the simulations.}
\label{contour-quint}
\end{figure}
\clearpage
\begin{figure}
\epsscale{1.00}
\plottwo{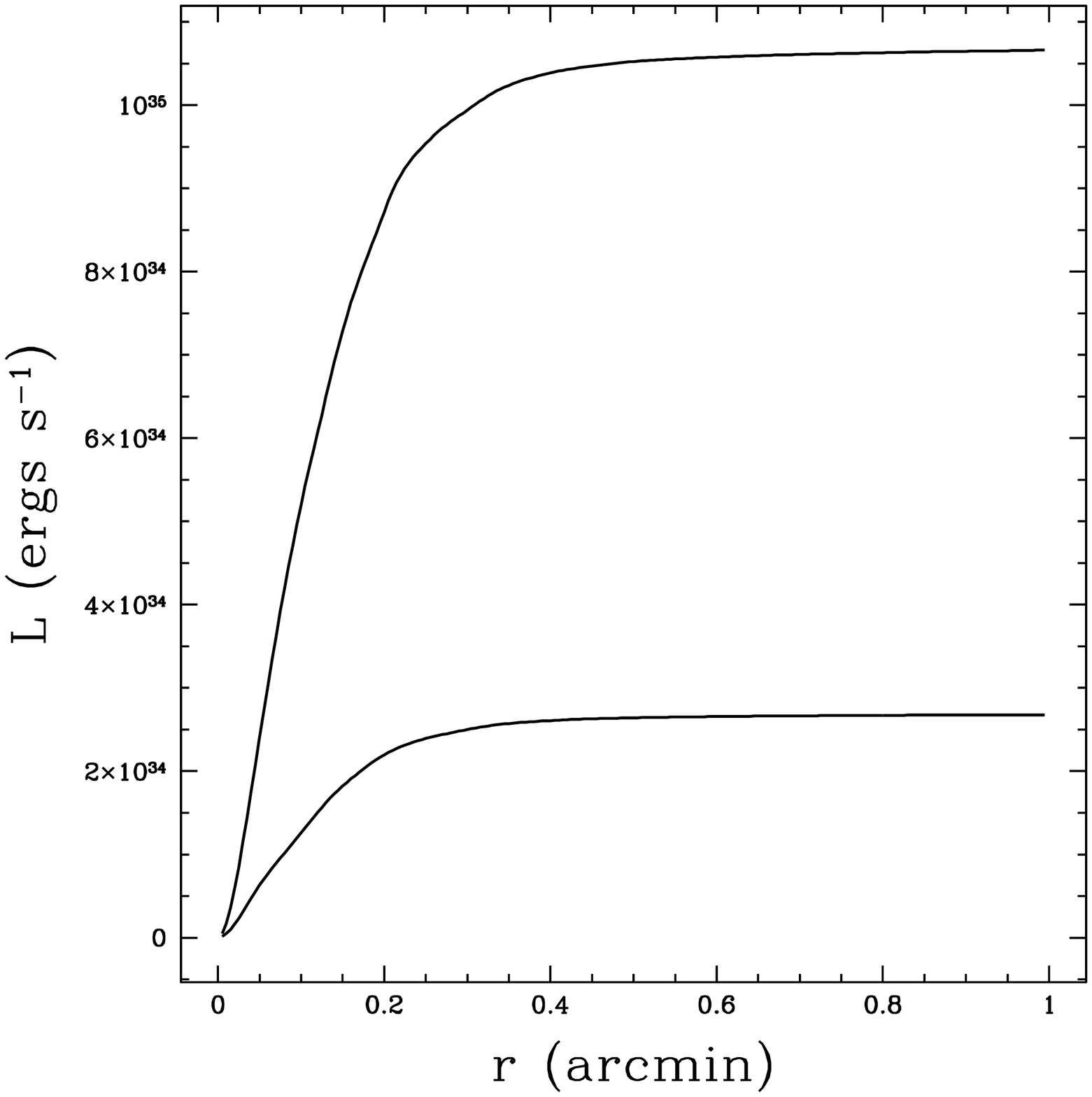}{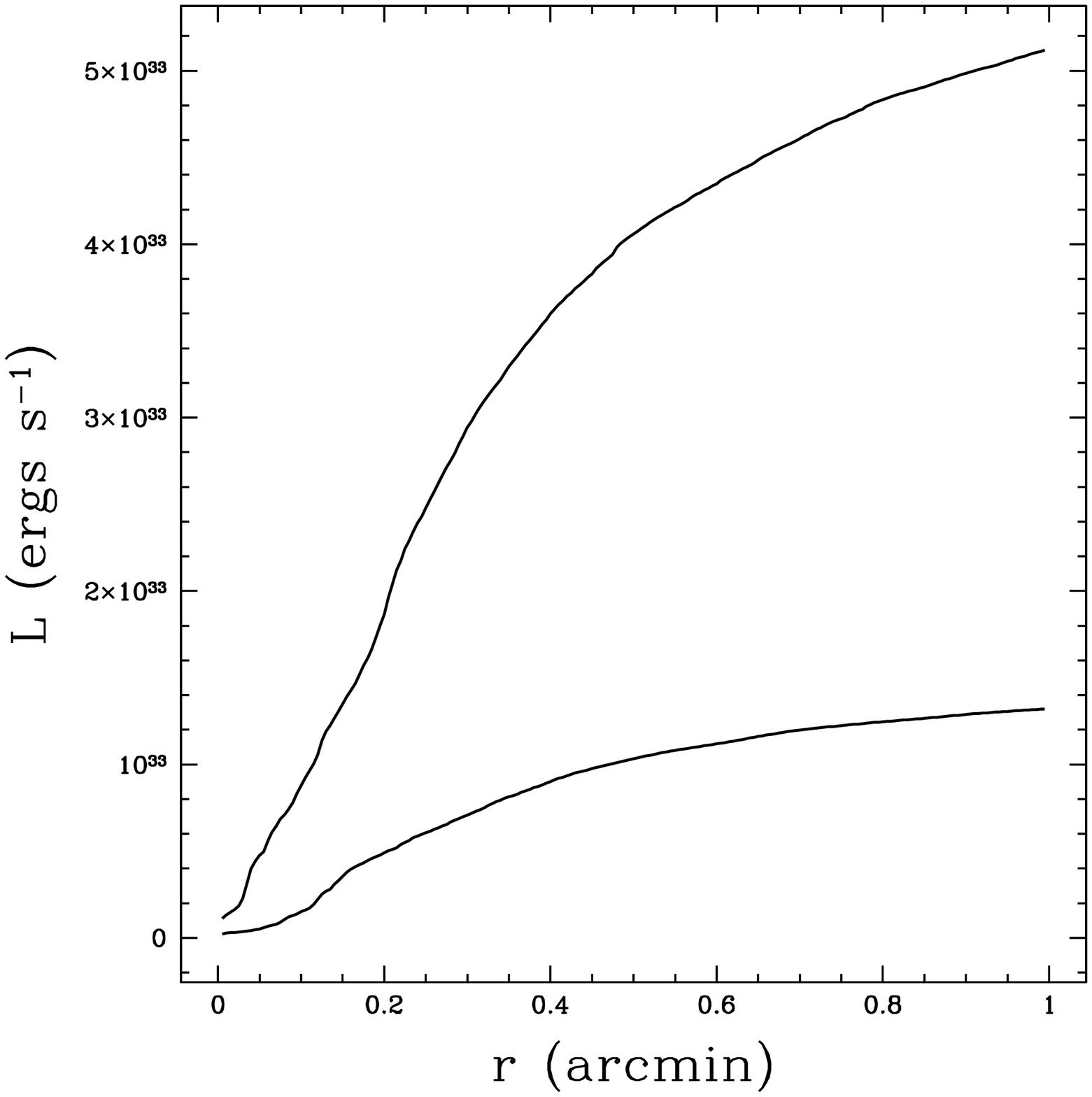}
\caption{Total 0.5--8~keV X-ray luminosity from concentric cylinders
  aligned with the center of the Arches (left) and Quintuplet
  clusters.  The lower line in each graph represents the ``standard''
  simulation of that cluster; the upper line represents the
  ``high-\mdot'' simulation.}
\label{lvsr}
\end{figure}
\clearpage
\begin{figure}
\epsscale{1.00}
\plottwo{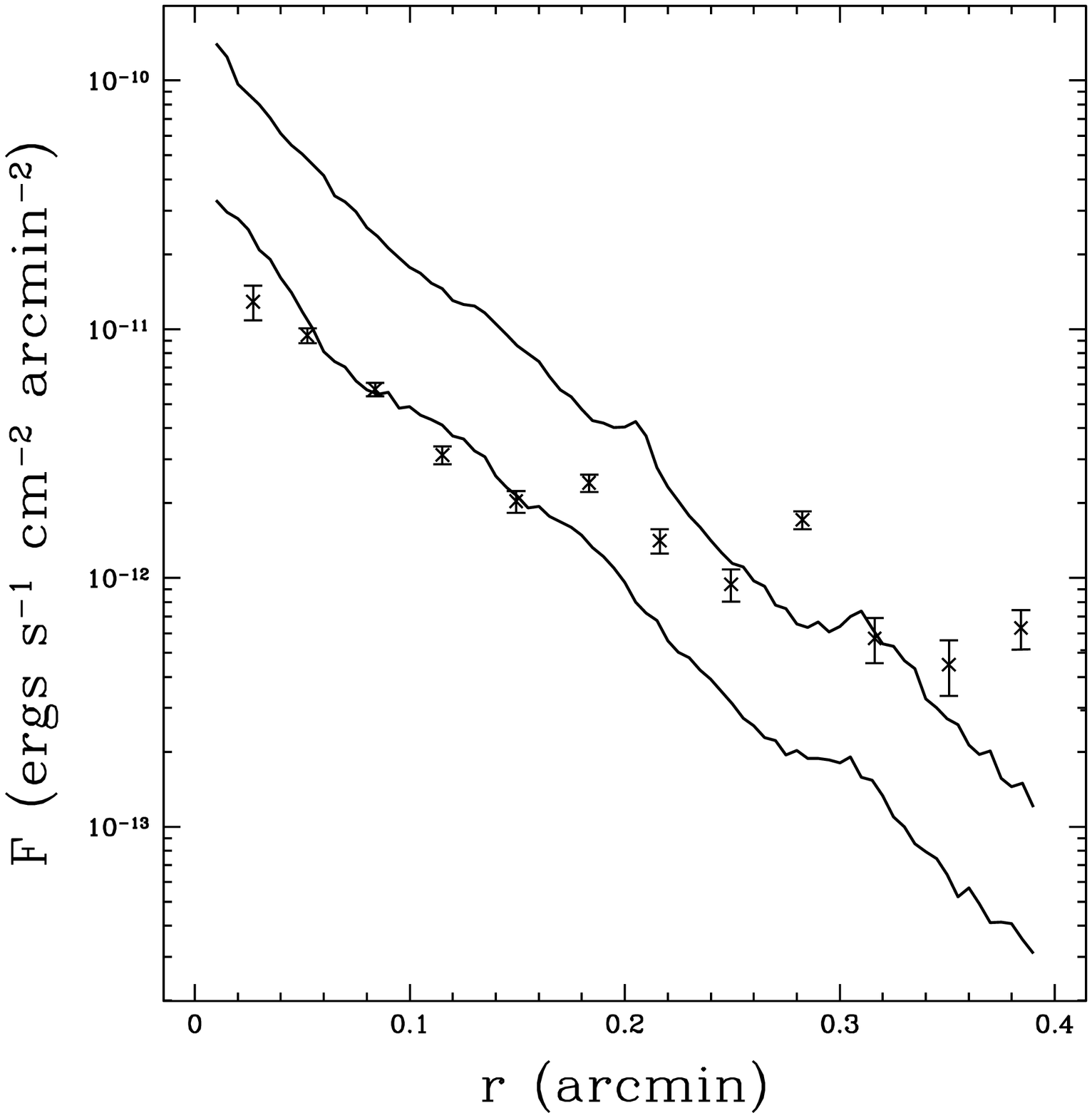}{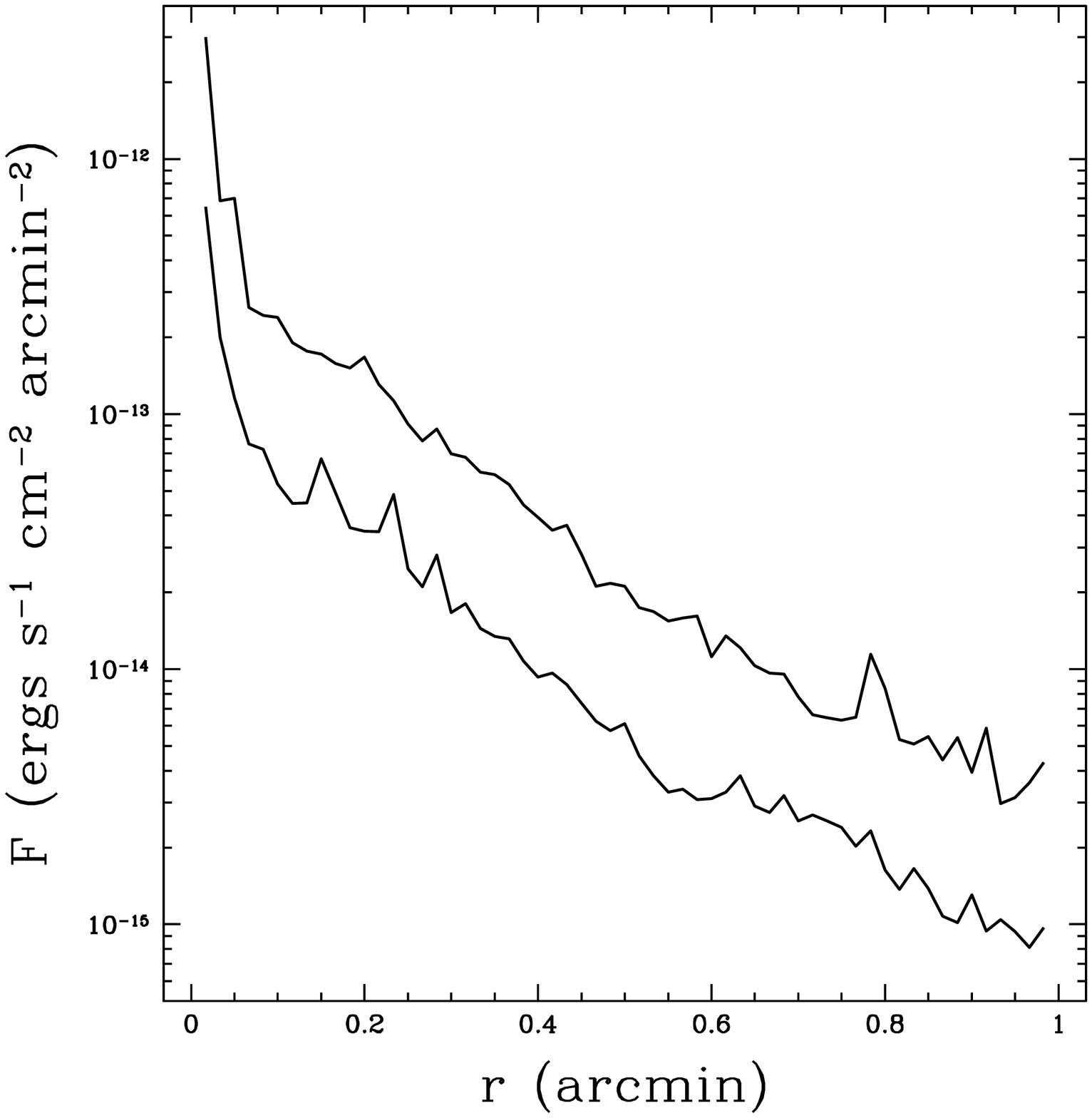}
\caption{The 2--8~keV X-ray flux per arcmin$^{2}$ from concentric
  cylinders aligned with the center of the Arches (left) and
  Quintuplet clusters.  The lower line in each graph represents the
  ``standard'' simulation of that cluster; the upper line represents
  the ``high-\mdot'' simulation.  Crosses and error bars in the left
  graph represent flux measurements from \textit{Chandra} observations
  of the Arches cluster.}
\label{fvsr}
\end{figure}
\clearpage
\begin{figure}
\epsscale{1.00}
\plottwo{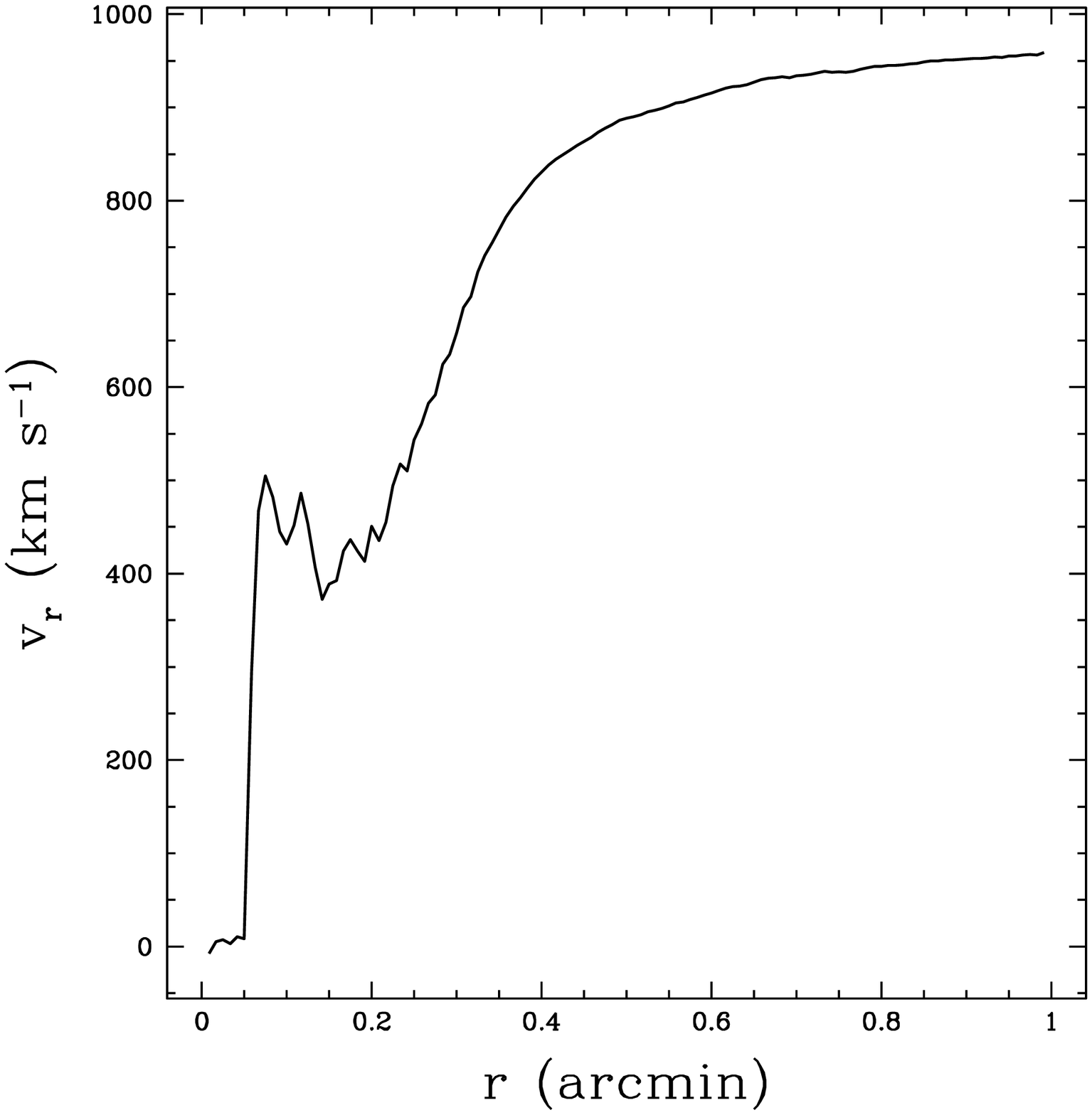}{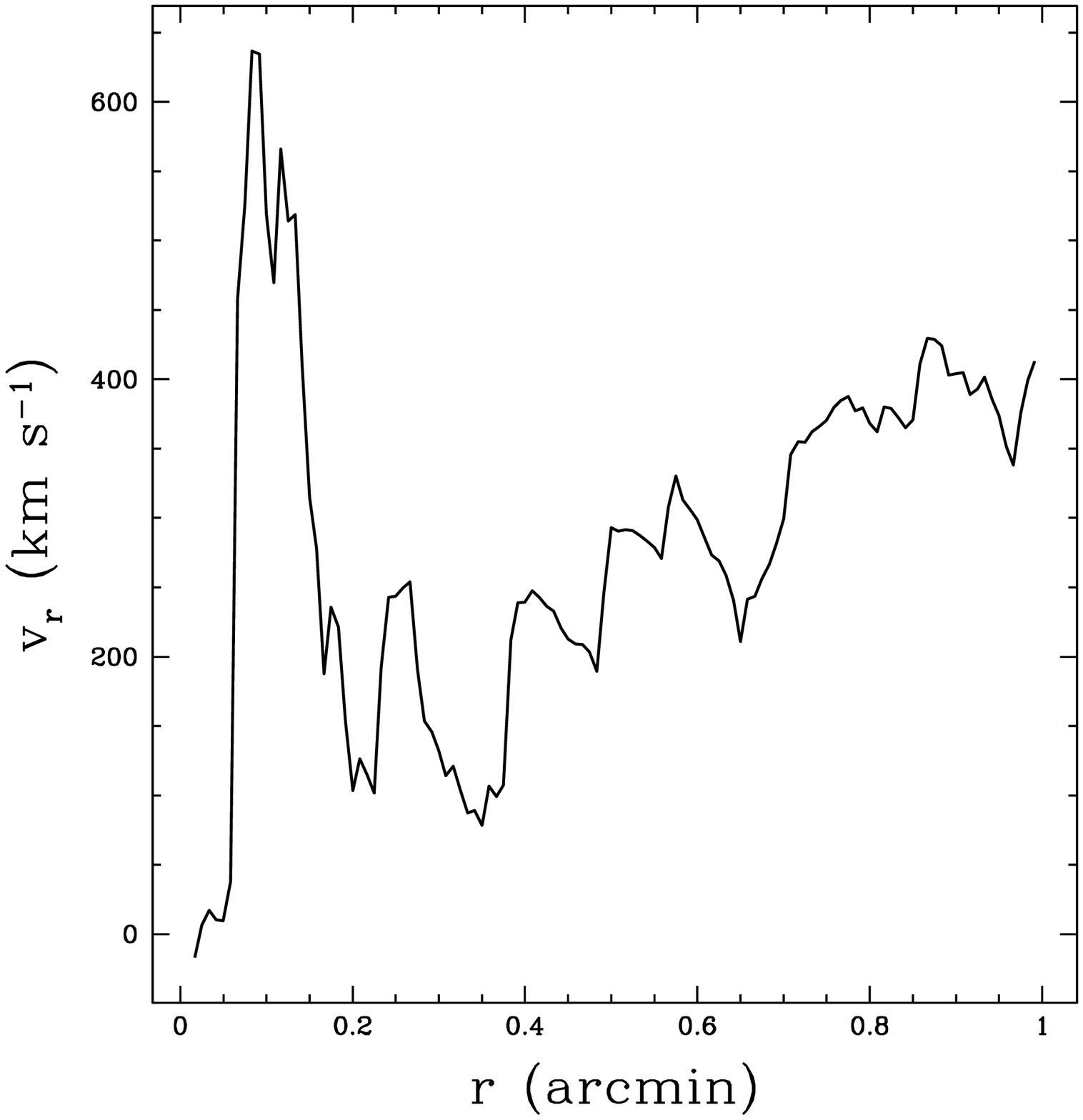}
\caption{Average radial velocity of gas in the ``standard'' simulations of 
  the Arches (left) and Quintuplet clusters, versus distance from the
  center of each cluster.}
\label{vvsr}
\end{figure}

\end{document}